\pdfoutput=1

\documentclass[11pt,twoside,a4paper,cmspaper,final,collab]{cms-tdr}
\def\svnVersion{245652}\def\cmsCernNo{2013-237}\def\cmsCernDate{\today}\def\cmsMessage{Published in Physical Review Letters as \href{http://dx.doi.org/10.1103/PhysRevLett.112.231802}{\doi{10.1103/PhysRevLett.112.231802.}}}

\begin{document}\cmsNoteHeader{TOP-12-040}

\hyphenation{had-ron-i-za-tion}
\hyphenation{cal-or-i-me-ter}
\hyphenation{de-vices}

\RCS$Revision: 238964 $
\RCS$HeadURL: svn+ssh://svn.cern.ch/reps/tdr2/papers/TOP-12-040/trunk/TOP-12-040.tex $
\RCS$Id: TOP-12-040.tex 238964 2014-04-29 14:06:29Z dnoonan $
\newlength\cmsFigWidth
\ifthenelse{\boolean{cms@external}}{\setlength\cmsFigWidth{0.85\columnwidth}}{\setlength\cmsFigWidth{0.4\textwidth}}
\ifthenelse{\boolean{cms@external}}{\providecommand{\cmsLeft}{top}}{\providecommand{\cmsLeft}{left}}
\ifthenelse{\boolean{cms@external}}{\providecommand{\cmsRight}{bottom}}{\providecommand{\cmsRight}{right}}

\newcommand\tW{\ensuremath{\cPqt\PW}\xspace}
\newcommand\DY{\ensuremath{\cPZ/\Pgg^{\star}}\xspace}
\newcommand\mtop{\ensuremath{m_{\cPqt}}\xspace}
\newcommand\ptsys{\ensuremath{\pt^{\text{sys}}}\xspace}
\newcommand{\Vvar}{\ensuremath{\cmsSymbolFace{V}}\xspace}
\ifthenelse{\boolean{cms@external}}{\providecommand{\CL}{\ensuremath{\mathrm{C.L.}}\xspace}}{\providecommand{\CL}{\ensuremath{\mathrm{CL}}\xspace}}
\cmsNoteHeader{TOP-12-040} % This is over-written in the CMS environment: useful as preprint no. for export versions
\title{Observation of the Associated Production of a Single Top Quark and a \texorpdfstring{\PW\ Boson in $\Pp\Pp$ Collisions at $\sqrt{s} = 8\TeV$}{W boson in pp collisions at sqrt(s)=8 TeV}}

\date{\today}

\abstract{
The first observation of the associated production of a single top quark and a $\PW$ boson is presented.
The analysis is based on a data set corresponding to an integrated luminosity of 12.2\fbinv of proton-proton collisions at $\sqrt{s} = 8\TeV$ recorded by the CMS experiment at the LHC.
Events with two leptons and a jet originating from a $\cPqb$ quark are selected.
A multivariate analysis based on kinematic and topological properties is used to separate the signal from the dominant \ttbar background.
An excess consistent with the signal hypothesis is observed, with a significance which corresponds to 6.1 standard deviations above a background-only hypothesis.
The measured production cross section is $23.4\pm5.4$\unit{pb}, in agreement with the standard model prediction.
}

\hypersetup{
pdfauthor={CMS Collaboration},
pdftitle={Observation of the associated production of a single top quark and a W boson in pp collisions at sqrt(s) = 8 TeV},
pdfsubject={CMS},
pdfkeywords={CMS, physics, top quark, single top}}

\maketitle %maketitle comes after all the front information has been supplied

Since its discovery in 1995 by the CDF \cite{Abe:1995hr} and D0 \cite{Abachi:1995iq} experiments at the Tevatron, studies of the top quark have raised great interest within high energy physics.
As the heaviest of all standard model (SM) particles, the top quark potentially plays an important role in electroweak symmetry breaking as well as in physics beyond the SM.
The measurement of the different mechanisms by which top quarks can be produced is instrumental in advancing the understanding of physics at the \TeV scale.

Top quarks are produced predominantly in pairs via the strong interaction in proton-proton ($\Pp\Pp$) collisions but they can also be produced singly via electroweak interactions, involving a $\PW\cPqt\cPqb$ vertex.
In the SM, single-top-quark production occurs mainly through three processes: $t$-channel ($\cPqt\cPq\cPqb$), $s$-channel ($\cPqt\cPqb$), and associated production of a top quark and a $\PW$ boson (\tW).
Single-top-quark production was first observed by the D0 \cite{Abazov:1165293} and CDF \cite{Aaltonen:1165362} experiments.
The $t$-channel production mode has been measured by D0 \cite{Abazov:2009pa,Abazov:2011rz} and CDF \cite{CDF-tchannel} as well as at the Large Hadron Collider (LHC) by the Compact Muon Solenoid (CMS) \cite{Chatrchyan:1359133} and ATLAS \cite{Aad:2012ux} experiments, while the observation of $s$-channel production was recently presented through a combination of the results of the CDF and D0 experiments \cite{CDF:2014uma}.
The \tW production cross section is negligible at the Tevatron, but large enough at the LHC to make it accessible.
Evidence for this process was presented by both the ATLAS \cite{Aad:2012dj} and CMS \cite{Chatrchyan:2012zca} experiments using the 7 \TeV collision data, with significances of 3.6 and $4.0\,\sigma$, respectively.
This Letter presents the first observation of \tW production at a significance of at least $5\,\sigma$, using data collected with the CMS experiment in $\Pp\Pp$ collisions at $\sqrt{s} = 8$ \TeV and corresponding to an integrated luminosity of 12.2\fbinv.

In addition to testing the SM predictions at the electroweak scale, associated \tW production is of interest because of its sensitivity to non-SM couplings of the $\PW\cPqt\cPqb$ vertex \cite{PhysRevD.63.014018, Cao200750, PhysRevD.81.034020, Godbole:2011vw, Ayazi:2013cba}, while being relatively insensitive to scenarios that affect the other single-top-quark production channels.

The theoretical prediction for the cross section of \tW production in $\Pp\Pp$ collisions at $\sqrt{s} = 8$\TeV at approximate next-to-next-to-leading order (NNLO) is $22.2 \pm 0.6\,\text{(scale)} \pm 1.4\,\mathrm{(PDF)}$\unit{pb} \cite{Kidonakis:2012rm}, with the first uncertainty coming from factorization and renormalization scale variations and the second from variations in the parton distribution functions of the proton.
At next-to-leading order (NLO), the definition of \tW production in perturbative quantum chromodynamics mixes with top-quark pair production (\ttbar) \cite{Frixione:2008yi,Belyaev:1998dn,White:2009yt}.
Two schemes for defining the \tW signal to distinguish it from \ttbar production have been proposed: the ``diagram removal'' (DR) \cite{Frixione:2008yi}, in which all doubly resonant NLO \tW diagrams are removed, and the ``diagram subtraction'' \cite{Frixione:2008yi, Tait:1999cf}, where a gauge-invariant subtraction term modifies the NLO \tW cross section to locally cancel the contribution from \ttbar.
In this Letter, the DR scheme is used for simulating the signal, but it was verified that the results are consistent between the two methods and any differences are accounted for in the systematic uncertainties.

The analysis is performed using the dilepton decay channels, in which the $\PW$ boson produced in association with the top quark and the $\PW$ boson from the decay of the top quark both decay leptonically into a muon or an electron, and a neutrino.
This leads to a final state composed of two oppositely charged isolated leptons, a jet resulting from the fragmentation of a $\cPqb$ quark, and two neutrinos.
The neutrinos escape detection and are only discernible by the presence of missing transverse energy (\MET), defined as the magnitude of the vector sum of the transverse momentum of all reconstructed particles.
The primary background to \tW production in this final state comes from \ttbar production, with \DY events being the next most significant.

The analysis uses a multivariate technique, exploiting kinematic and topological differences to distinguish the \tW signal from the dominant \ttbar background.
To assess the robustness of the result, two additional analyses were conducted.
One involves a fit to a single kinematic variable, the other is based on event counts.

The central feature of the CMS apparatus \cite{Chatrchyan:2008zzk} is a superconducting solenoid with an internal diameter of 6 \unit{m}, providing a magnetic field of 3.8 \unit{T}.
Within the bore of the solenoid are a silicon pixel and strip tracker, a lead tungstate crystal electromagnetic calorimeter, and a brass/scintillator hadron calorimeter.
Muons are measured in gas-ionization detectors embedded in the steel flux return yoke outside the magnet.
In addition, CMS has extensive forward calorimetry.
The detector covers a region of $\abs{\eta} < 5.0$, where the pseudorapidity $\eta$ is defined as $\eta = - \ln [\tan(\theta/2)]$, where $\theta$ is the polar angle.

Data samples are selected based on triggers requiring two leptons (either an electron or muon), one with transverse momentum, \pt, of at least 17\GeV and a second with \pt of at least 8\GeV.
All events are required to have a well-reconstructed primary vertex \cite{CMS-PAS-TRK-10-005}.
The primary vertex with the largest sum of $\pt^2$ of associated tracks is chosen.

Electrons are reconstructed from energy deposits in the electromagnetic calorimeter (including energy deposits from radiated photons) matched to tracks in the silicon tracker.
Muons, \MET, and jets are reconstructed using the CMS particle flow (PF) algorithm \cite{particleFlow1, particleFlow2}, which performs a global event reconstruction.
Electrons and muons are required to have $\pt > 20$\GeV and fall within the pseudorapidity range of $\abs{\eta} < 2.5$ for electrons and $\abs{\eta}< 2.4$ for muons.
Exactly two oppositely charged, isolated leptons are required in the event, and events are rejected if they contain additional leptons passing a looser criteria, for which the \pt threshold is lowered to 10\GeV.
In order to limit the contribution from low-mass dilepton resonances, the invariant mass of the dilepton system, $m_{\ell\ell}$ $(\ell = \Pe$ or $\mu)$, is required to be greater than 20\GeV.
Events in the $\Pe\Pe$ and $\mu\mu$ final states are rejected if $m_{\ell\ell}$ is between 81 and 101\GeV, to suppress the $\cPZ \rightarrow \ell \ell$ process in the same-flavor final states.
Additionally, a requirement of $\MET > 50$\GeV is applied for these final states.

Jets are reconstructed by clustering PF candidates using the anti-\kt algorithm \cite{antiKt} with a distance parameter of 0.5.
Selected jets must be within $\abs{\eta} < 2.4$ and have $\pt > 30$\GeV.
Corrections are made to the jet energies for detector response as a function of $\eta$ and $\pt$ \cite{Chatrchyan:2011ds}.
Additional corrections are made to subtract energy in the jet from multiple $\Pp\Pp$ collisions (pileup) \cite{Soyez:2012hv}.
Jets originating from the decay of a $\cPqb$ quark are tagged based on the presence of a secondary vertex, identified using a multivariate algorithm combining tracking information in a discriminant \cite{Chatrchyan:2012jua}.
A working point is chosen, corresponding to a $\cPqb$-tagging efficiency of approximately 70\% and with a misidentification rate of 1\%--2\%.
Loose jets, whose discrimination power against \ttbar background is discussed later, are defined as jets failing the requirements on \pt and $\eta$, but passing the less restrictive selection requirement of $\pt > 20$\GeV and $\abs{\eta} < 4.9$, while still passing all other selection criteria. In particular, loose jets that fall within $\abs{\eta} < 2.4$ are classified as central loose jets.

For events passing the dilepton and \MET criteria described above, a region in which the \tW signal is enhanced (signal region) and two regions dominated by background (control regions) are defined.
The signal region contains events with exactly one jet passing the selection requirements, which is $\cPqb$ tagged ($1j1t$ region).
Two control regions enriched in \ttbar background are defined as having exactly two jets with either one or both being $\cPqb$ tagged ($2j1t$ and $2j2t$ regions, respectively).

Events from Monte Carlo simulation are used to estimate the contributions and kinematics of signal and background processes.
Single-top-quark events are simulated at NLO with the \POWHEG 1.0 event generator \cite{Nason:2004rx, Frixione:2007vw, Alioli:2010xd,Re:2010bp}; \MADGRAPH 5.1.3 is used for simulating \ttbar and single-boson events ($\Vvar$+jets, where $\Vvar=\PW,\cPZ$) \cite{madgraph5}.
Samples are produced using a top-quark mass \mtop$=172.5$\GeV, consistent with its current best measurement \cite{Aaltonen:2012ra}.
Diboson backgrounds are simulated using \PYTHIA 6.426 \cite{pythia}.
In all samples, fragmentation and hadronization are modeled with \PYTHIA, and \TAUOLA v27.121.5 is used to simulate $\tau$ decays \cite{tauola}.
The CTEQ6L1 and CTEQ6.6M PDF sets \cite{Nadolsky:2008zw} are used for samples simulated at leading-order and NLO, respectively.
A full simulation of the response of the CMS detector is performed for all generated events using a \GEANTfour-based model \cite{Geant4NIM}.
The simulation includes modeling of pileup, with the distribution of the number of interactions in simulation matching that in data.
Simulated samples are normalized to the NNLO cross sections for \ttbar [$\sigma_{\ttbar} = 245.8^{+6.2}_{-8.4}\,(\text{scale})^{+6.2}_{-6.4}\,\text{(PDF)}$\unit{pb}] \cite{Czakon:2013goa}, \DY, and $\PW$+jets processes, with approximate NNLO cross sections used for single top quark \cite{Kidonakis:2012rm} and NLO for diboson processes.
The \DY simulation is reweighted to reproduce the \MET distribution observed in data, using events with $m_{\ell\ell}$ in the vicinity of the \cPZ-boson mass (81 to 101\GeV) to derive scale factors.

After the selection, the simulated samples in the $1j1t$ signal region contain predominantly \tW and \ttbar events (comprising 16\% and 76\% of the events, respectively), with a smaller contribution from \DY events (6\%).
The two control regions are dominated by \ttbar production.
Event yields in simulation and data in the signal and control regions are shown in Table \ref{tab:eventCounts}.

\begin{table*}[t]
\topcaption{Event yields in the signal and control regions. Yields from simulation are shown with statistical (first) and systematic (second) uncertainties.}
\label{tab:eventCounts}
\centering
\setlength{\extrarowheight}{0.5ex}
\begin{scotch}{lccc}
                 &  $1j1t$                         & $2j1t$                        & $2j2t$ \\
\hline
\tW              &  1500$\pm$20$\pm$130   & 790$\pm$20$\pm$80       & 220$\pm$10$\pm$30 \\
\ttbar           &  7090$\pm$60$\pm$900   & 12\,910$\pm$80$\pm$1320   & 7650$\pm$60$\pm$1020 \\
\DY, other     &  670$\pm$30$\pm$90     & 370$\pm$30$\pm$60       & 36$\pm$7$\pm$12  \\
\hline
Total simulation &  9260$\pm$70$\pm$1040  & 14\,070$\pm$90$\pm$1410   & 7910$\pm$70$\pm$1020 \\
Data             &  9353              & 13\,479               & 7615 \\
\end{scotch}
\end{table*}

\begin{figure}
    \begin{center}
      \includegraphics[width=0.4\textwidth]{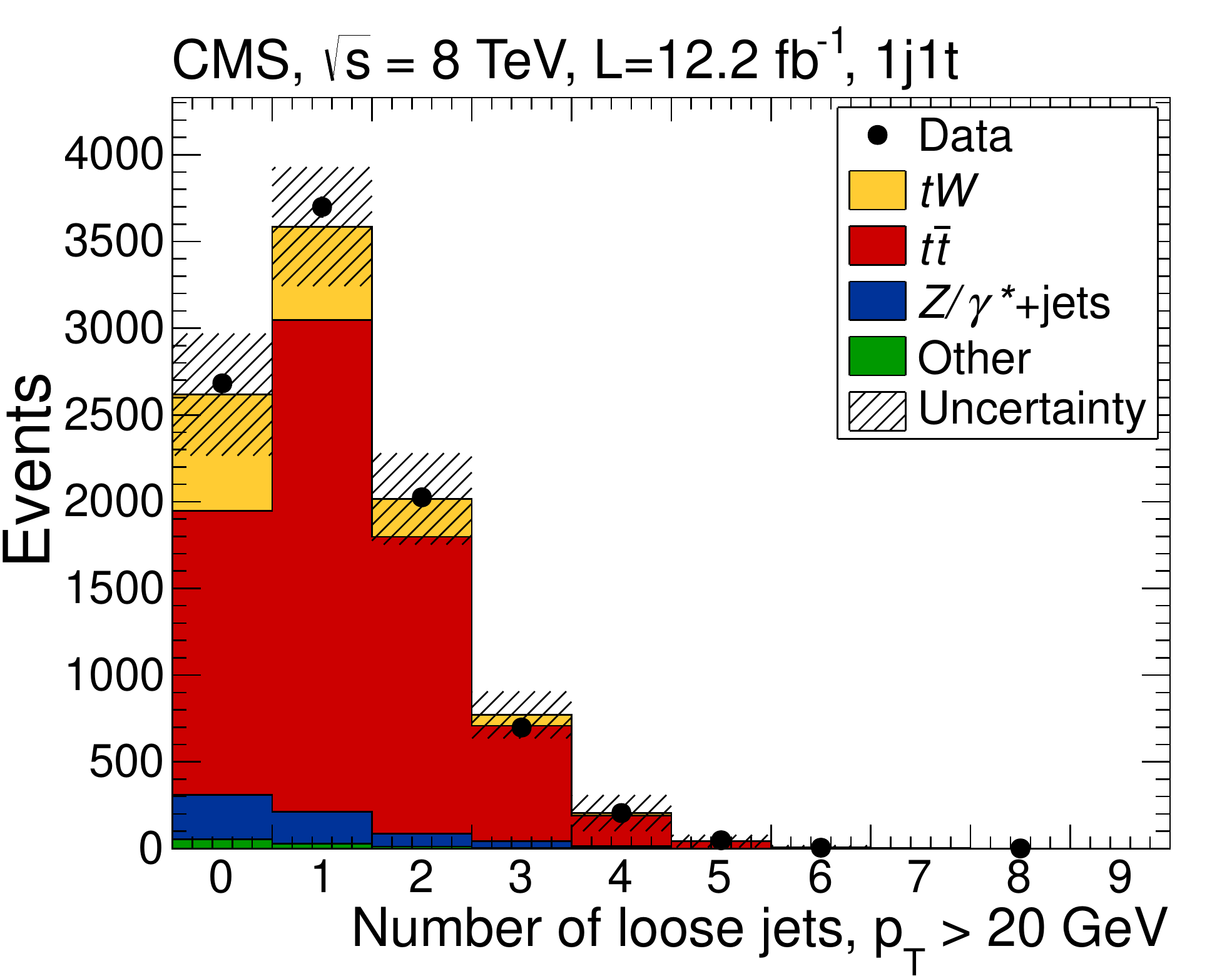}
      \includegraphics[width=0.4\textwidth]{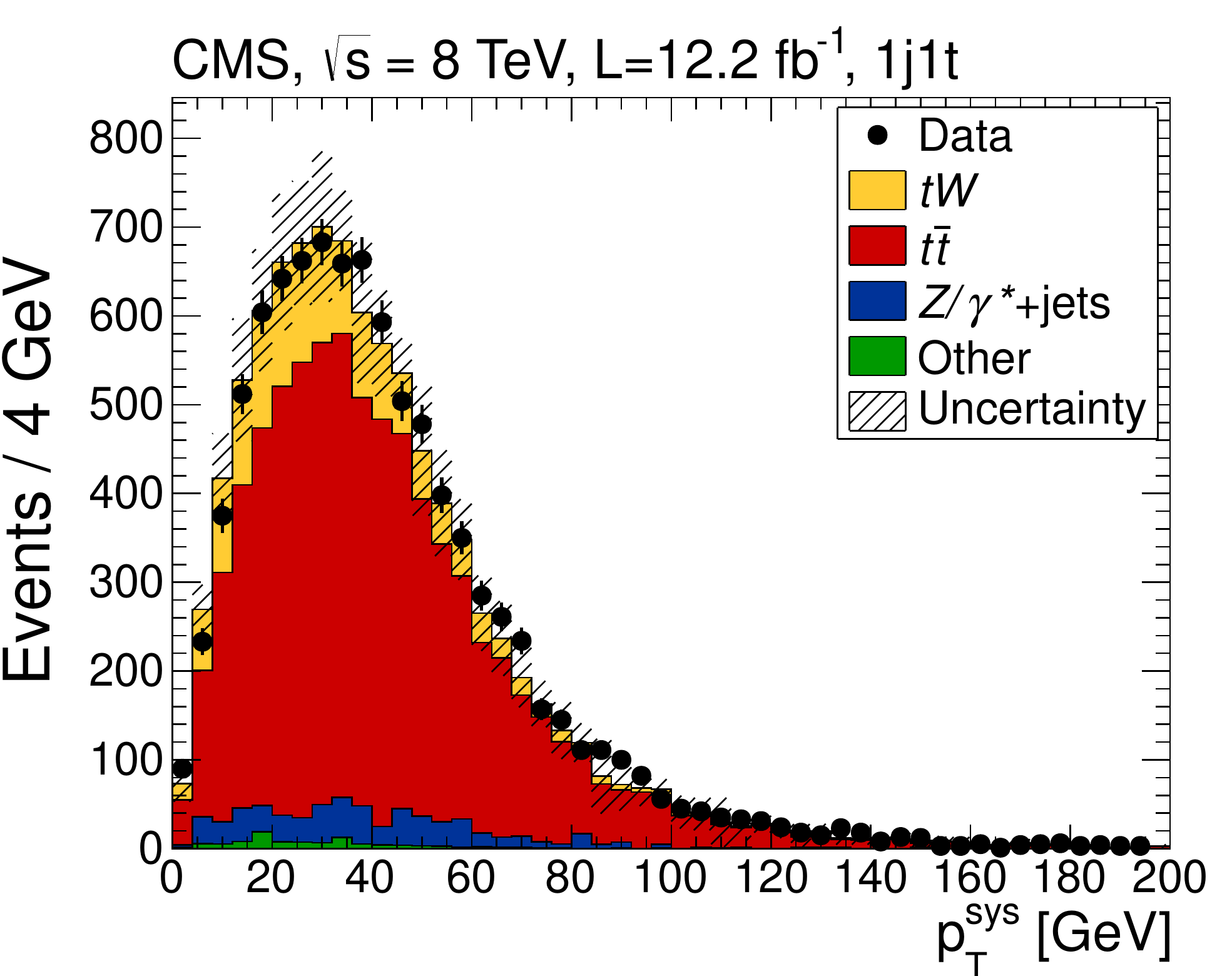}
      \caption{\label{fig:nloosejetptsys}  The number of loose jets in the event and the \pt of the system (\ptsys) composed of the jet, leptons, and \MET, in the signal region ($1j1t$) for all final states combined. Shown are data (points) and simulation (histogram). The hatched band represents the combined effect of all sources of systematic uncertainty.}
    \end{center}
  \end{figure}

In order to separate the \tW signal from the \ttbar background, a multivariate analysis based on boosted decision trees (BDT) \cite{decision_trees} is used, implemented with the toolkit for multivariate data analysis \cite{tmva}.
The BDT analyzer is trained using 13 variables, chosen for their separation power in distinguishing \tW and \ttbar, as well as being well modeled in simulation when checked in control regions.
The most powerful variables are those involving loose jets in the event: the number of loose jets, number of central loose jets, and the number of loose jets that are $\cPqb$ tagged.
Other variables with significant separation power are related to the kinematics of the system comprised of the leptons, jets and \MET: the scalar sum of their transverse momenta ($\HT$), the magnitude of the vector sum of their transverse momenta (\ptsys), and invariant mass of the system.
A complete list of the variables used can be found in the Supplemental Material \ifthenelse{\boolean{cms@external}}{\cite{Supplemental}}{(Appendix \ref{app:suppMat})}.
The distributions of the number of loose jets and the \pt of the system in the $1j1t$ signal region are shown in Fig.~\ref{fig:nloosejetptsys} for all three final states ($\Pe\Pe$, $\Pe\mu$, and $\mu\mu$) combined.

\begin{figure}
    \begin{center}
      \includegraphics[width=0.4\textwidth]{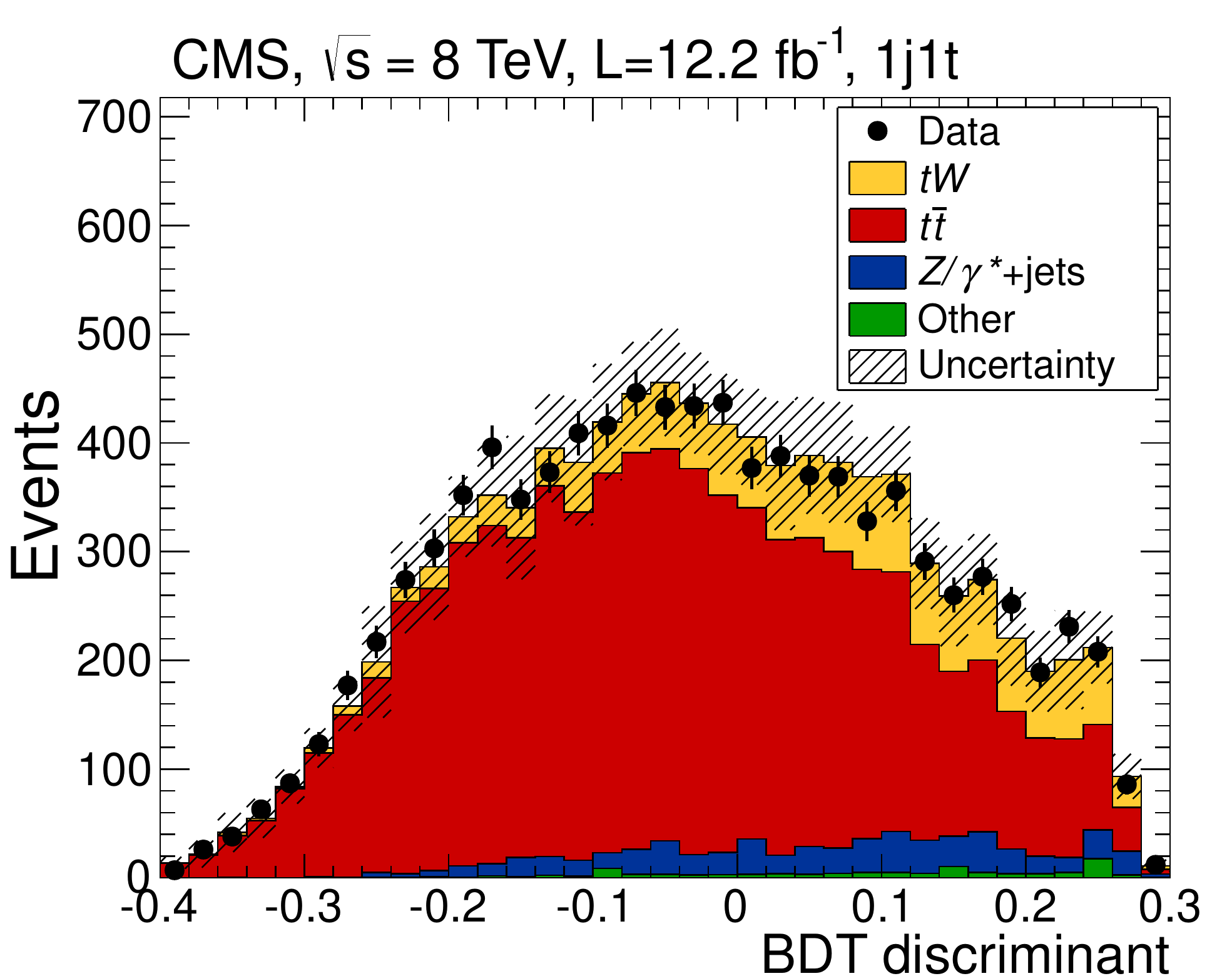}\\
      \includegraphics[width=0.4\textwidth]{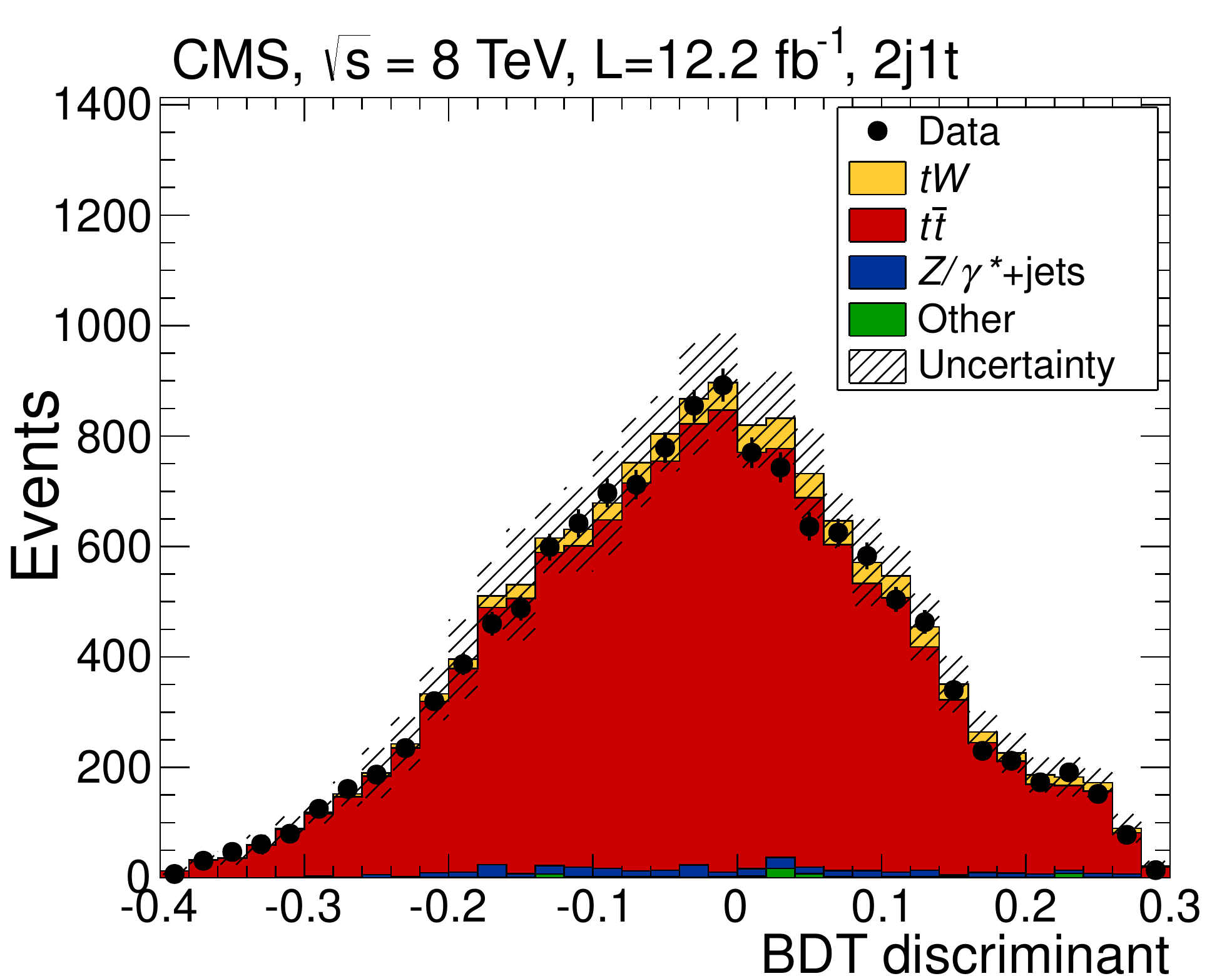}
      \includegraphics[width=0.4\textwidth]{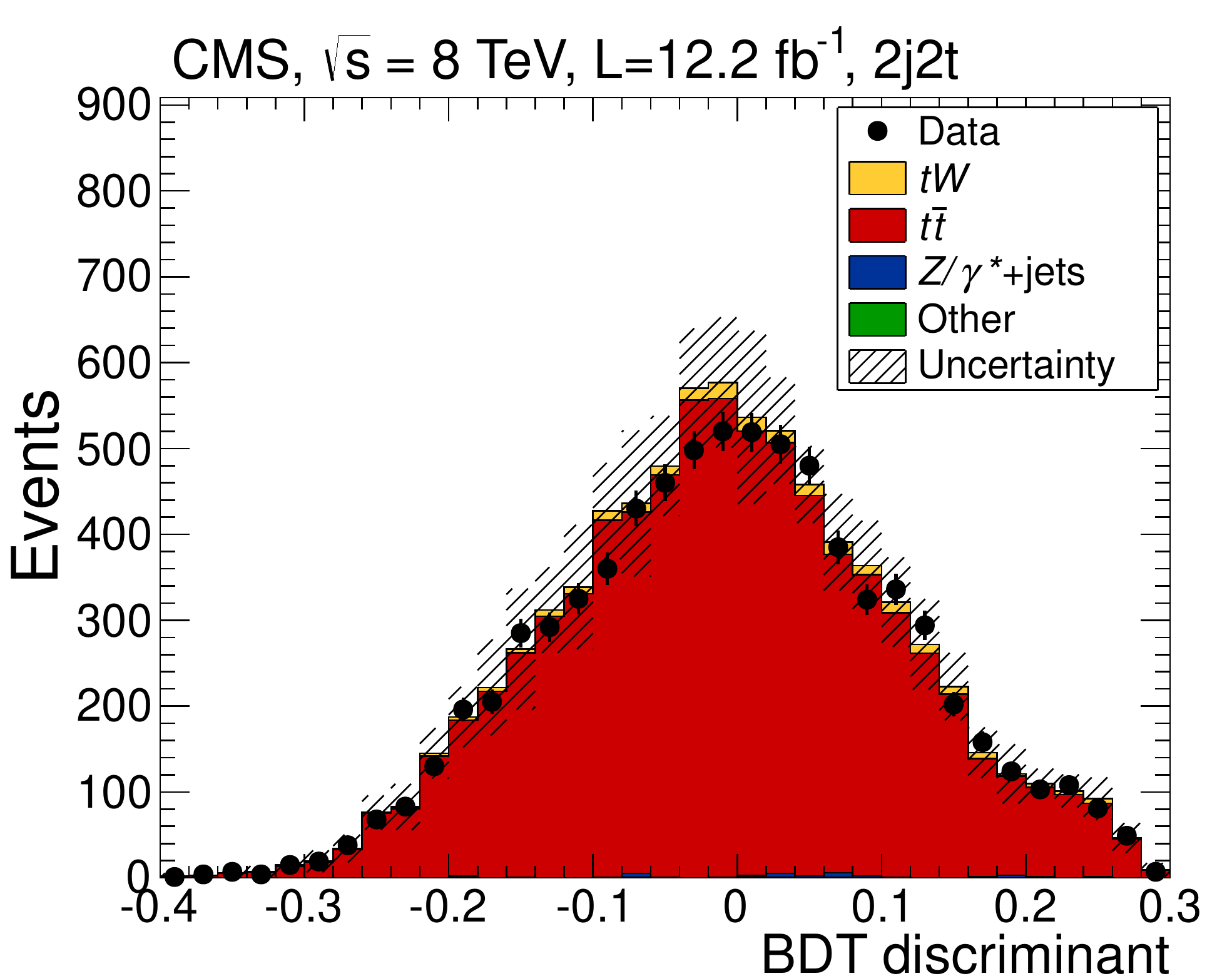}
      \caption{\label{fig:bdt}  The BDT discriminant, in the signal region ($1j1t$) and control regions ($2j1t$ and $2j2t$) for all final states combined. Shown are data (points) and simulation (histogram). The hatched band represents the combined effect of all sources of systematic uncertainty.}
    \end{center}
  \end{figure}

The BDT analyzer provides a single discriminant value for each event.
The distributions of the BDT discriminant in data and simulation are shown in Fig.~\ref{fig:bdt} for the $1j1t$, $2j1t$, and $2j2t$ regions, combining all three final states together.

The uncertainty from all systematic sources is determined by estimating their effect on the normalization and shape of the BDT discriminant for all regions and final states.
The dominant systematic uncertainties come from the choice of thresholds for the matrix element and parton showering (ME/PS) matching in simulation of \ttbar production and the renormalization/factor\-i\-za\-tion scale.
The effect of these uncertainties was estimated by producing simulated samples with the value of the ME/PS matching thresholds and renormalization/factor\-i\-za\-tion scale doubled and halved from their respective initial values of 20\GeV and $\mtop^2 + \sum\pt^2$ (where the sum is over all additional final state partons), contributing a 14\% and 12\% uncertainty, respectively, to the measured cross section.
The uncertainty due to the value of the top-quark mass used in simulation is estimated by simulating \tW and \ttbar processes with a varied value for \mtop, resulting in a 9\% effect on the cross section.
The complete list of systematic uncertainties and corresponding effects on the cross section can be found in the Supplemental Material \ifthenelse{\boolean{cms@external}}{\cite{Supplemental}}{(Appendix \ref{app:suppMat})}.

A simultaneous binned likelihood fit to the rate and shape of the BDT distributions of the three final states in the three regions is performed.
The two control regions are included in the fit to allow for better determination of the \ttbar contribution.
The distributions for signal and background are taken from simulation.
In the likelihood function, for each source of systematic uncertainty \cPqu, a nuisance parameter $\theta_\cPqu$ is introduced.
The rates of signal and background are allowed to vary in the fit, constrained in the likelihood function by the systematic uncertainties.
The excess of events is quantified based on the score statistic $q$, chosen to enhance numerical stability, defined as
\begin{equation*}
q = \frac{\partial}{\partial\mu}\ln\mathcal{L}(\mu = 0, \hat{\theta}_{0} | \text{data}),
\end{equation*}
where $\mu$ is the signal strength parameter (defined as the signal cross section in units of the SM prediction) and $\hat{\theta}_{0}$ is the set of nuisance parameters that maximizes the likelihood $\mathcal{L}$ for a background-only hypothesis ($\mu = 0$).
The score statistic is evaluated for sets of four billion pseudoexperiments using a background-only hypothesis.
The significance is determined based on the probability of producing a score statistic value in the background-only hypothesis as high or higher than that observed in data.
The expected significance is evaluated using the median and central 68\% interval of the score statistic values obtained in pseudo-experiments generated under a signal-plus-background hypothesis.
A profile likelihood method is used to determine the signal cross section and 68\% confidence level (\CL) interval.

We observe an excess of events above the expected background with a $p$ value of $5\times10^{-10}$ corresponding to a significance of $6.1\,\sigma$, compared to an expected significance from simulation of $5.4\pm1.4\,\sigma$.
The measured cross section is found to be $23.4\pm5.4$\unit{pb}, where the uncertainty is mainly systematic, in agreement with the predicted SM value of $22.2 \pm 0.6\,\text{(scale)} \pm 1.4\,\text{(PDF)}$\unit{pb}.

The cross section measurement is used to determine the absolute value of the Cabibbo-Koba\-ya\-shi-Maskawa matrix element $\abs{V_{\cPqt\cPqb}}$, assuming $\abs{V_{\cPqt\cPqb}} \gg \abs{V_{\cPqt\cPqd}}$ and $\abs{V_{\cPqt\cPqs}}$
\begin{equation*}
\abs{V_{\cPqt\cPqb}} = \sqrt{\sigma_{\cPqt\PW} / \sigma^{\text{th}}_{\cPqt\PW}} = 1.03 \pm 0.12\,\text{(exp)} \pm 0.04\,\text{(th.)},
\end{equation*}
where $\sigma^\text{th}_{\cPqt\PW}$ is the theoretical prediction of the \tW cross section assuming $\abs{V_{\cPqt\cPqb}} = 1$, and the uncertainties are separated into experimental and theoretical values.
Using the SM assumption $ 0 \le \abs{V_{\cPqt\cPqb}}^2 \le 1 $, a lower bound $\abs{V_{\cPqt\cPqb}} > 0.78$ at 95\% \CL is found using the approach of Feldman and Cousins \cite{FeldmanCousins}.

\begin{figure}
    \begin{center}
      \includegraphics[width=0.4\textwidth]{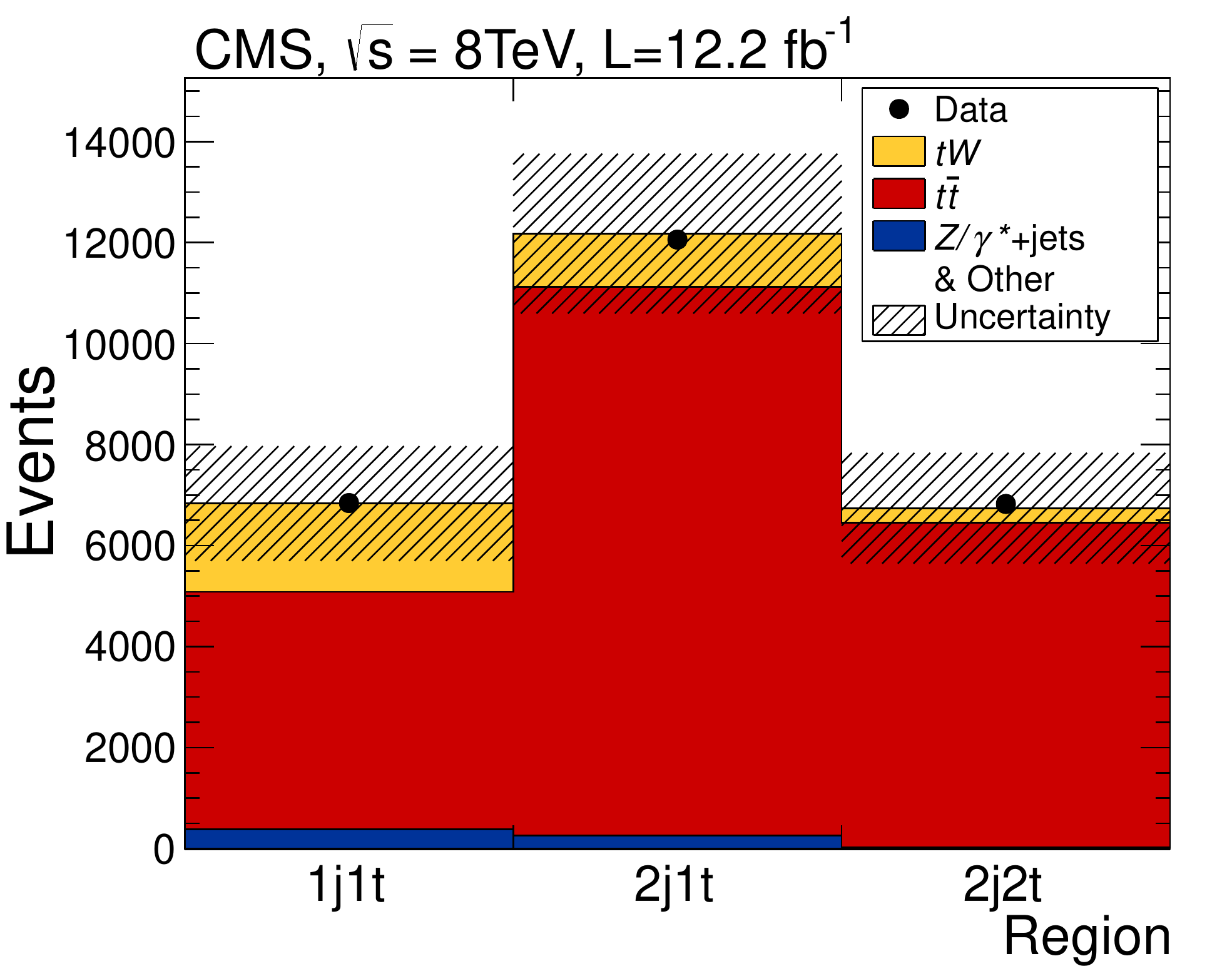}
      \caption{\label{fig:crosscheck}  Event yields in data and simulation for events passing additional requirements from the cross-check analyses.  Yields are shown in the $1j1t$ signal regions and $2j1t$ and $2j2t$ control regions for a combination of all three final states, with the simulation scaled to the outcome of the statistical fit from the event-count analysis.  The hatched band represents the combined effect of all systematic uncertainties on the event yields.}
    \end{center}
  \end{figure}

Using the same selection as in the BDT analysis, two cross-check analyses are performed.
Events containing any $\cPqb$ tagged loose jets are rejected.
Additionally, a requirement of $\HT > 160$\GeV is added in the $\Pe\mu$ final state, where no \MET requirement is applied.
The effects of systematic uncertainties are taken into account in the same way as for the BDT analysis, and the same method for extraction of the significance and cross section is used.
The first cross-check analysis is based on the distribution of \ptsys rather than the BDT discriminant, and results in an observed significance of $4.0\,\sigma$ above a background-only hypothesis, with an expected significance of $3.2^{+0.4}_{-0.9}\,\sigma$, and a measured cross section of $24.3\pm8.6$\unit{pb}.
The second cross-check analysis is based only on event counts after selection, and an excess of events is observed above the background with a significance of $3.6\,\sigma$, with an expected significance based on simulation of $2.8\pm0.9\,\sigma$, and a measured cross section of $33.9\pm8.6$\unit{pb}.
Event yields in data and simulation for this analysis are shown in Fig.~\ref{fig:crosscheck}, with the simulation scaled to the result of the statistical fit.
The results of both analyses are consistent with those found in the BDT analysis, but with larger, mostly systematic, uncertainties.

In summary, the production of a single top quark in association with a $\PW$ boson is observed for the first time.
The analysis uses data collected by the CMS experiment in $\Pp\Pp$ collisions at $\sqrt{s} = 8$\TeV, corresponding to an integrated luminosity of 12.2\fbinv.
An excess of events above background is found with a significance of $6.1\,\sigma$, and a \tW production cross section of $23.4\pm5.4$\unit{pb} is measured, in agreement with the standard model prediction.

We congratulate our colleagues in the CERN accelerator departments for the excellent performance of the LHC and thank the technical and administrative staffs at CERN and at other CMS institutes for their contributions to the success of the CMS effort. In addition, we gratefully acknowledge the computing centers and personnel of the Worldwide LHC Computing Grid for delivering so effectively the computing infrastructure essential to our analyses. Finally, we acknowledge the enduring support for the construction and operation of the LHC and the CMS detector provided by the following funding agencies: BMWF and FWF (Austria); FNRS and FWO (Belgium); CNPq, CAPES, FAPERJ, and FAPESP (Brazil); MES (Bulgaria); CERN; CAS, MoST, and NSFC (China); COLCIENCIAS (Colombia); MSES and CSF (Croatia); RPF (Cyprus); MoER, SF0690030s09 and ERDF (Estonia); Academy of Finland, MEC, and HIP (Finland); CEA and CNRS/IN2P3 (France); BMBF, DFG, and HGF (Germany); GSRT (Greece); OTKA and NIH (Hungary); DAE and DST (India); IPM (Iran); SFI (Ireland); INFN (Italy); NRF and WCU (Republic of Korea); LAS (Lithuania); CINVESTAV, CONACYT, SEP, and UASLP-FAI (Mexico); MBIE (New Zealand); PAEC (Pakistan); MSHE and NSC (Poland); FCT (Portugal); JINR (Dubna); MON, RosAtom, RAS and RFBR (Russia); MESTD (Serbia); SEIDI and CPAN (Spain); Swiss Funding Agencies (Switzerland); NSC (Taipei); ThEPCenter, IPST, STAR and NSTDA (Thailand); TUBITAK and TAEK (Turkey); NASU (Ukraine); STFC (United Kingdom); DOE and NSF (USA).

\bibliography{auto_generated}   % will be created by the tdr script.

\providecommand{\href}[2]{#2}\begingroup\raggedright\begin{thebibliography}{10}%
\makeatletter
\providecommand{\hrefCMSnoop }[0]{\@secondoftwo}%
\makeatother
\providecommand{\doi}{\texttt{doi:}\begingroup \urlstyle{tt}\Url}

\bibitem{Abe:1995hr}
\hrefCMSnoop {} {{ CDF} Collaboration, ``{Observation of top quark production
  in $\Pap\Pp$ collisions with the Collider Detector at Fermilab}'',} \textit{
  Phys. Rev. Lett.} \textbf{ 74} (1995) 2626,
\href{http://dx.doi.org/10.1103/PhysRevLett.74.2626}{\doi{10.1103/PhysRevLett.74.2626}}.
%%CITATION = HEP-EX/9503002;%%.

\bibitem{Abachi:1995iq}
\hrefCMSnoop {} {{ D0} Collaboration, ``{Observation of the top quark}'',}
  \textit{ Phys. Rev. Lett.} \textbf{ 74} (1995) 2632,
\href{http://dx.doi.org/10.1103/PhysRevLett.74.2632}{\doi{10.1103/PhysRevLett.74.2632}}.
%%CITATION = HEP-EX/9503003;%%.

\bibitem{Abazov:1165293}
\hrefCMSnoop {} {{ D0} Collaboration, ``{Observation of Single Top-Quark
  Production}'',} \textit{ Phys. Rev. Lett.} \textbf{ 103} (2009) 092001,
\href{http://dx.doi.org/10.1103/PhysRevLett.103.092001}{\doi{10.1103/PhysRevLett.103.092001}}.
%%CITATION = 0903.0850;%%.

\bibitem{Aaltonen:1165362}
\hrefCMSnoop {} {{ CDF} Collaboration, ``{Observation of Electroweak Single Top
  Quark Production}'',} \textit{ Phys. Rev. Lett.} \textbf{ 103} (2009) 092002,
  \href{http://dx.doi.org/10.1103/PhysRevLett.103.092002}{\doi{10.1103/PhysRevLett.103.092002}}.

\bibitem{Abazov:2009pa}
\hrefCMSnoop {} {{ D0} Collaboration, ``{Measurement of the t-channel single
  top quark production cross section}'',} \textit{ Phys. Lett. B} \textbf{ 682}
  (2010) 363,
\href{http://dx.doi.org/10.1016/j.physletb.2009.11.038}{\doi{10.1016/j.physletb.2009.11.038}}.
%%CITATION = ARXIV:0907.4259;%%.

\bibitem{Abazov:2011rz}
\hrefCMSnoop {} {{ D0} Collaboration, ``{Model-independent measurement of
  $t$-channel single top quark production in $\Pp\Pap$ collisions at
  $\sqrt{s}=1.96$ TeV}'',} \textit{ Phys. Lett. B} \textbf{ 705} (2011) 313,
\href{http://dx.doi.org/10.1016/j.physletb.2011.10.035}{\doi{10.1016/j.physletb.2011.10.035}}.
%%CITATION = ARXIV:1105.2788;%%.

\bibitem{CDF-tchannel}
\hrefCMSnoop {} {{ CDF} Collaboration, ``{Observation of Single Top Quark
  Production and Measurement of $|V_{tb}|$ with CDF}'',} \textit{ Phys. Rev. D}
  \textbf{ 82} (2010) 112005,
\href{http://dx.doi.org/10.1103/PhysRevD.82.112005}{\doi{10.1103/PhysRevD.82.112005}}.
%%CITATION = ARXIV:1004.1181;%%.

\bibitem{Chatrchyan:1359133}
\hrefCMSnoop {} {{ CMS} Collaboration, ``{Measurement of the t-channel single
  top quark production cross section in pp collisions at $\sqrt{s}$ = 7
  TeV}'',} \textit{ Phys. Rev. Lett.} \textbf{ 107} (2011) 091802,
  \href{http://dx.doi.org/10.1103/PhysRevLett.107.091802}{\doi{10.1103/PhysRevLett.107.091802}}.

\bibitem{Aad:2012ux}
\hrefCMSnoop {} {{ ATLAS} Collaboration, ``{Measurement of the t-channel single
  top-quark production cross section in pp collisions at $\sqrt{s} = 7$ TeV
  with the ATLAS detector}'',} \textit{ Phys. Lett. B} \textbf{ 717} (2012)
  330,
  \href{http://dx.doi.org/10.1016/j.physletb.2012.09.031}{\doi{10.1016/j.physletb.2012.09.031}}.

\bibitem{CDF:2014uma}
\hrefCMSnoop {} {{CDF and D0 Collaborations}, ``{Observation of s-channel
  production of single top quarks at the Tevatron}'',} (2014).
\href{http://www.arXiv.org/abs/1402.5126}{\texttt{ arXiv:1402.5126}}.
%%CITATION = ARXIV:1402.5126;%%.

\bibitem{Aad:2012dj}
\hrefCMSnoop {} {{ ATLAS} Collaboration, ``{Evidence for the associated
  production of a W boson and a top quark in ATLAS at $\sqrt{s}$ = 7 TeV}'',}
  \textit{ Phys. Lett. B} \textbf{ 716} (2012) 142,
\href{http://dx.doi.org/10.1016/j.physletb.2012.08.011}{\doi{10.1016/j.physletb.2012.08.011}}.
%%CITATION = ARXIV:1205.5764;%%.

\bibitem{Chatrchyan:2012zca}
\hrefCMSnoop {} {{ CMS} Collaboration, ``{Evidence for associated production of
  a single top quark and W boson in pp collisions at $\sqrt{s}$ = 7 TeV}'',}
  \textit{ Phys. Rev. Lett.} \textbf{ 110} (2013) 022003,
\href{http://dx.doi.org/10.1103/PhysRevLett.110.022003}{\doi{10.1103/PhysRevLett.110.022003}}.
%%CITATION = ARXIV:1209.3489;%%.

\bibitem{PhysRevD.63.014018}
\hrefCMSnoop {} {T.~M.~P. Tait and C.-P. Yuan, ``Single top quark production as
  a window to physics beyond the standard model'',} \textit{ Phys. Rev. D}
  \textbf{ 63} (2000) 014018,
  \href{http://dx.doi.org/10.1103/PhysRevD.63.014018}{\doi{10.1103/PhysRevD.63.014018}}.

\bibitem{Cao200750}
\hrefCMSnoop {} {Q.-H. Cao, J.~Wudka, and C.-P. Yuan, ``Search for new physics
  via single-top production at the LHC'',} \textit{ Phys. Lett. B} \textbf{
  658} (2007) 50,
  \href{http://dx.doi.org/10.1016/j.physletb.2007.10.057}{\doi{10.1016/j.physletb.2007.10.057}}.

\bibitem{PhysRevD.81.034020}
\hrefCMSnoop {} {V.~Barger, M.~McCaskey, and G.~Shaughnessy, ``Single top and
  Higgs associated production at the LHC'',} \textit{ Phys. Rev. D} \textbf{
  81} (2010) 034020,
  \href{http://dx.doi.org/10.1103/PhysRevD.81.034020}{\doi{10.1103/PhysRevD.81.034020}}.

\bibitem{Godbole:2011vw}
\hrefCMSnoop {} {R.~M. Godbole, L.~Hartgring, I.~Niessen, and C.~D. White,
  ``{Top polarisation studies in $H^-t$ and $Wt$ production}'',} \textit{ JHEP}
  \textbf{ 01} (2012) 011,
\href{http://dx.doi.org/10.1007/JHEP01(2012)011}{\doi{10.1007/JHEP01(2012)011}}.
%%CITATION = ARXIV:1111.0759;%%.

\bibitem{Ayazi:2013cba}
\hrefCMSnoop {} {S.~Y. Ayazi, H.~Hesari, and M.~M. Najafabadi, ``{Probing the
  top quark chromoelectric and chromomagnetic dipole moments in single top
  $tW$-channel at the LHC}'',} \textit{ Phys.Lett.} \textbf{ B727} (2013)
  199--203,
  \href{http://dx.doi.org/10.1016/j.physletb.2013.10.025}{\doi{10.1016/j.physletb.2013.10.025}},
\href{http://www.arXiv.org/abs/1307.1846}{\texttt{ arXiv:1307.1846}}.
%%CITATION = ARXIV:1307.1846;%%.

\bibitem{Kidonakis:2012rm}
\hrefCMSnoop {} {N.~Kidonakis, ``{NNLL threshold resummation for top-pair and
  single-top production}'',} (2012).
\href{http://www.arXiv.org/abs/1210.7813}{\texttt{ arXiv:1210.7813}}.
%%CITATION = ARXIV:1210.7813;%%.

\bibitem{Frixione:2008yi}
S.~Frixione\hrefCMSnoop {} { {et~al.}, ``{Single-top hadroproduction in
  association with a W boson}'',} \textit{ JHEP} \textbf{ 07} (2008) 029,
  \href{http://dx.doi.org/10.1088/1126-6708/2008/07/029}{\doi{10.1088/1126-6708/2008/07/029}}.

\bibitem{Belyaev:1998dn}
\hrefCMSnoop {} {A.~S. Belyaev, E.~E. Boos, and L.~V. Dudko, ``{Single top
  quark at future hadron colliders: Complete signal and background study}'',}
  \textit{ Phys. Rev. D} \textbf{ 59} (1999) 075001,
\href{http://dx.doi.org/10.1103/PhysRevD.59.075001}{\doi{10.1103/PhysRevD.59.075001}}.
%%CITATION = HEP-PH/9806332;%%.

\bibitem{White:2009yt}
\hrefCMSnoop {} {C.~D. White, S.~Frixione, E.~Laenen, and F.~Maltoni,
  ``{Isolating Wt production at the LHC}'',} \textit{ JHEP} \textbf{ 11} (2009)
  074,
  \href{http://dx.doi.org/10.1088/1126-6708/2009/11/074}{\doi{10.1088/1126-6708/2009/11/074}},
  \href{http://www.arXiv.org/abs/0908.0631}{\texttt{ arXiv:0908.0631}}.

\bibitem{Tait:1999cf}
\hrefCMSnoop {} {T.~M.~P. Tait, ``{The $tW^-$ mode of single top
  production}'',} \textit{ Phys. Rev. D} \textbf{ 61} (1999) 034001,
\href{http://dx.doi.org/10.1103/PhysRevD.61.034001}{\doi{10.1103/PhysRevD.61.034001}}.
%%CITATION = HEP-PH/9909352;%%.

\bibitem{Chatrchyan:2008zzk}
\hrefCMSnoop {} {{ CMS} Collaboration, ``The {CMS} experiment at the {CERN}
  {LHC}'',} \textit{ JINST} \textbf{ 03} (2008) S08004,
\href{http://dx.doi.org/10.1088/1748-0221/3/08/S08004}{\doi{10.1088/1748-0221/3/08/S08004}}.
%%CITATION = JINST,3,S08004;%%.

\bibitem{CMS-PAS-TRK-10-005}
\href {http://cdsweb.cern.ch/record/1279383} {{ CMS} Collaboration, ``Tracking
  and Primary Vertex Results in First 7 TeV Collisions'',} CMS Physics Analysis
  Summary CMS-PAS-TRK-10-005, 2010.

\bibitem{particleFlow1}
\href {http://cdsweb.cern.ch/record/1194487} {{ CMS} Collaboration, ``Particle
  Flow Event Reconstruction in CMS and Performance for Jets, Taus, and MET'',}
  CMS Physics Analysis Summary CMS-PAS-PFT-09-001, 2009.

\bibitem{particleFlow2}
\href {http://cdsweb.cern.ch/record/1279341} {{ CMS} Collaboration,
  ``Commissioning of the Particle-Flow Reconstruction in Minimum-Bias and Jet
  Events from pp Collisions at 7 TeV'',} CMS Physics Analysis Summary
  CMS-PAS-PFT-10-002, 2010.

\bibitem{antiKt}
\hrefCMSnoop {} {M.~Cacciari, G.~P. Salam, and G.~Soyez, ``{The anti-$k_t$ jet
  clustering algorithm}'',} \textit{ JHEP} \textbf{ 04} (2008) 063,
  \href{http://dx.doi.org/10.1088/1126-6708/2008/04/063}{\doi{10.1088/1126-6708/2008/04/063}}.

\bibitem{Chatrchyan:2011ds}
\hrefCMSnoop {} {{ CMS} Collaboration, ``{Determination of Jet Energy
  Calibration and Transverse Momentum Resolution in CMS}'',} \textit{ JINST}
  \textbf{ 6} (2011) P11002,
  \href{http://dx.doi.org/10.1088/1748-0221/6/11/P11002}{\doi{10.1088/1748-0221/6/11/P11002}},
\href{http://www.arXiv.org/abs/1107.4277}{\texttt{ arXiv:1107.4277}}.
%%CITATION = ARXIV:1107.4277;%%.

\bibitem{Soyez:2012hv}
G.~Soyez\hrefCMSnoop {} { {et~al.}, ``{Pileup subtraction for jet shapes}'',}
  \textit{ Phys. Rev. Lett.} \textbf{ 110} (2013) 162001,
  \href{http://dx.doi.org/10.1103/PhysRevLett.110.162001}{\doi{10.1103/PhysRevLett.110.162001}},
\href{http://www.arXiv.org/abs/1211.2811}{\texttt{ arXiv:1211.2811}}.
%%CITATION = ARXIV:1211.2811;%%.

\bibitem{Chatrchyan:2012jua}
\hrefCMSnoop {} {{ CMS} Collaboration, ``{Identification of b-quark jets with
  the CMS experiment}'',} \textit{ JINST} \textbf{ 8} (2013) P04013,
\href{http://dx.doi.org/10.1088/1748-0221/8/04/P04013}{\doi{10.1088/1748-0221/8/04/P04013}}.
%%CITATION = ARXIV:1211.4462;%%.

\bibitem{Nason:2004rx}
\hrefCMSnoop {} {P.~Nason, ``{A New method for combining NLO QCD with shower
  Monte Carlo algorithms}'',} \textit{ JHEP} \textbf{ 11} (2004) 040,
  \href{http://dx.doi.org/10.1088/1126-6708/2004/11/040}{\doi{10.1088/1126-6708/2004/11/040}},
\href{http://www.arXiv.org/abs/hep-ph/0409146}{\texttt{ arXiv:hep-ph/0409146}}.
%%CITATION = HEP-PH/0409146;%%.

\bibitem{Frixione:2007vw}
\hrefCMSnoop {} {S.~Frixione, P.~Nason, and C.~Oleari, ``{Matching NLO QCD
  computations with Parton Shower simulations: the POWHEG method}'',} \textit{
  JHEP} \textbf{ 11} (2007) 070,
  \href{http://dx.doi.org/10.1088/1126-6708/2007/11/070}{\doi{10.1088/1126-6708/2007/11/070}},
\href{http://www.arXiv.org/abs/0709.2092}{\texttt{ arXiv:0709.2092}}.
%%CITATION = ARXIV:0709.2092;%%.

\bibitem{Alioli:2010xd}
\hrefCMSnoop {} {S.~Alioli, P.~Nason, C.~Oleari, and E.~Re, ``{A general
  framework for implementing NLO calculations in shower Monte Carlo programs:
  the POWHEG BOX}'',} \textit{ JHEP} \textbf{ 06} (2010) 043,
  \href{http://dx.doi.org/10.1007/JHEP06(2010)043}{\doi{10.1007/JHEP06(2010)043}},
\href{http://www.arXiv.org/abs/1002.2581}{\texttt{ arXiv:1002.2581}}.
%%CITATION = ARXIV:1002.2581;%%.

\bibitem{Re:2010bp}
\hrefCMSnoop {} {E.~Re, ``{Single-top $Wt$-channel production matched with
  parton showers using the POWHEG method}'',} \textit{ Eur. Phys. J. C}
  \textbf{ 71} (2011) 1547,
  \href{http://dx.doi.org/10.1140/epjc/s10052-011-1547-z}{\doi{10.1140/epjc/s10052-011-1547-z}},
\href{http://www.arXiv.org/abs/1009.2450}{\texttt{ arXiv:1009.2450}}.
%%CITATION = ARXIV:1009.2450;%%.

\bibitem{madgraph5}
J.~Alwall\hrefCMSnoop {} { {et~al.}, ``{MadGraph} 5: going beyond'',} \textit{
  JHEP} \textbf{ 06} (2011) 128,
\href{http://dx.doi.org/10.1007/JHEP06(2011)128}{\doi{10.1007/JHEP06(2011)128}}.
%%CITATION = ARXIV:1106.0522;%%.

\bibitem{Aaltonen:2012ra}
\hrefCMSnoop {} {{CDF and D0 Collaborations}, ``{Combination of the top-quark
  mass measurements from the Tevatron collider}'',} \textit{ Phys. Rev. D}
  \textbf{ 86} (2012) 092003,
  \href{http://dx.doi.org/10.1103/PhysRevD.86.092003}{\doi{10.1103/PhysRevD.86.092003}},
  \href{http://www.arXiv.org/abs/1207.1069}{\texttt{ arXiv:1207.1069}}.
An update can be found in
  \href{http://www.arXiv.org/abs/arXiv:1305.3929}{arXiv:1305.3929}.
%%CITATION = ARXIV:1207.1069;%%.

\bibitem{pythia}
\hrefCMSnoop {} {T.~Sj{\"o}strand, S.~Mrenna, and P.~Skands, ``{PYTHIA} 6.4
  physics and manual'',} \textit{ JHEP} \textbf{ 05} (2006) 026,
\href{http://dx.doi.org/10.1088/1126-6708/2006/05/026}{\doi{10.1088/1126-6708/2006/05/026}}.
%%CITATION = HEP-PH/0603175;%%.

\bibitem{tauola}
N.~Davidson\hrefCMSnoop {} { {et~al.}, ``Universal interface of {TAUOLA}
  technical and physics documentation'',} \textit{ Comput. Phys. Commun.}
  \textbf{ 183} (2012) 821,
\href{http://dx.doi.org/10.1016/j.cpc.2011.12.009}{\doi{10.1016/j.cpc.2011.12.009}}.
%%CITATION = ARXIV:1002.0543;%%.

\bibitem{Nadolsky:2008zw}
P.~M. Nadolsky\hrefCMSnoop {} { {et~al.}, ``{Implications of CTEQ global
  analysis for collider observables}'',} \textit{ Phys. Rev. D} \textbf{ 78}
  (2008) 013004,
\href{http://dx.doi.org/10.1103/PhysRevD.78.013004}{\doi{10.1103/PhysRevD.78.013004}}.
%%CITATION = ARXIV:0802.0007;%%.

\bibitem{Geant4NIM}
\hrefCMSnoop {} {{ GEANT4} Collaboration, ``{GEANT4}---a simulation toolkit'',}
  \textit{ Nucl. Instrum. Meth. A} \textbf{ 506} (2003) 250,
\href{http://dx.doi.org/10.1016/S0168-9002(03)01368-8}{\doi{10.1016/S0168-9002(03)01368-8}}.
%%CITATION = NUIMA,A506,250;%%.

\bibitem{Czakon:2013goa}
\hrefCMSnoop {} {M.~Czakon, P.~Fiedler, and A.~Mitov, ``{The total top quark
  pair production cross-section at hadron colliders through
  O($\alpha_S^4$)}'',} \textit{ Phys. Rev. Lett.} \textbf{ 110} (2013) 252004,
\href{http://dx.doi.org/10.1103/PhysRevLett.110.252004}{\doi{10.1103/PhysRevLett.110.252004}}.
%%CITATION = ARXIV:1303.6254;%%.

\bibitem{decision_trees}
L.~Breiman, J.~Friedman, C.~J. Stone, and R.~A. Olshen, ``Classification and
  Regression Trees''.
\newblock Chapman and Hall, 1984.

\bibitem{tmva}
\href {http://pos.sissa.it/archive/conferences/050/040/ACAT_040.pdf} {H.~Voss,
  A.~H{\"o}cker, J.~Stelzer, and F.~Tegenfeldt, ``{TMVA}, the Toolkit for
  Multivariate Data Analysis with {ROOT}'',} in \textit{ XI International
  Workshop on Advanced Computing and Analysis Techniques in Physics Research},
  p.~040.
\newblock SISSA, 2007.
\newblock \href{http://www.arXiv.org/abs/physics/0703039}{\texttt{
  arXiv:physics/0703039}}.
\newblock
{PoS(ACAT2007)040}.
%%CITATION = PHYSICS/0703039;%%.

\bibitem{FeldmanCousins}
\hrefCMSnoop {} {G.~J. Feldman and R.~D. Cousins, ``A unified approach to the
  classical statistical analysis of small signals'',} \textit{ Phys. Rev. D}
  \textbf{ 57} (1998) 3873,
  \href{http://dx.doi.org/10.1103/PhysRevD.57.3873}{\doi{10.1103/PhysRevD.57.3873}},
\href{http://www.arXiv.org/abs/physics/9711021}{\texttt{
  arXiv:physics/9711021}}.
%%CITATION = ARXIV:1105.2788;%%.

\end{thebibliography}\endgroup

\ifthenelse{\boolean{cms@external}}{}{
\clearpage
\appendix
\section{Supplemental Material\label{app:suppMat}}
Two additional tables are provided giving details on the variables used in the BDT analysis and additional information on the sources of systematic uncertainty.
Table \ref{tab:inputVariables} gives the definition of the thirteen variables used as inputs to the training of the BDT analyzer.
Table \ref{tab:syst} presents the contribution of each of the sources of systematic uncertainty in the analysis to the measured cross section uncertainty, as well as a brief description of the systematic uncertainty.
The contribution due to each source are estimated by fixing the nuisance parameters associated with each of the sources one at a time, and measuring the effect on the uncertainty on the cross section.

Two additional sets of figures are provided.
Figure \ref{fig:inputVariables} shows the distribution of the eleven additional input variables used in the analysis.
Figure \ref{fig:bdt0jets} shows the distribution of the BDT in two additional control regions, containing events with either exactly one jet which fails the \cPqb-tagging requirements (1j0t region) or two jets which both fail the \cPqb-tagging requirements (2j0t).
These two control regions were used to validate the shape of the BDT discriminant in the two largest backgrounds, \ttbar and \DY.

Event yields in the signal and control regions, split into the three dilepton final states, are shown in Tables \ref{tab:eventCountsemu}, \ref{tab:eventCountsmumu}, and \ref{tab:eventCountsee} for the $\Pe\Pe$, $\Pe\mu$, $\mu\mu$, final states, respectively.

\begin{table*}[hb]
\topcaption{Variables used for BDT training}
\label{tab:inputVariables}
\centering
\resizebox{\textwidth}{!}{
\begin{scotch}{ l l }
Variable Name & Description \\
\hline
\# of loose jets                         &  Number of loose jets, $\pt >20$\GeV, $\abs{\eta} < 4.9$ \\
\# of central loose jets                 &  Number of loose jets, $\pt >20$\GeV, $\abs{\eta} <  2.4$ \\
\# of b-tagged loose jets                &  Number of loose jets, $\pt >20$\GeV, \Pqb-tagged, $\abs{\eta} < 2.4$ \\
\ptsys                                   &  Vector sum of $\pt$ of leptons, jet, and \MET \\
$\HT$                                    &  Scalar sum of $\pt$ of leptons, jet, and \MET \\
$\pt(\text{jet})$                        &  \pt of the leading, tight, b-tagged jet \\
$\pt(\text{loose jet})$                  &  \pt of leading loose jet, defined as 0 for events with no loose jet present \\
\ptsys/\HT                               &  Ratio of \ptsys to \HT for the event \\
$m_{\text{sys}}$                         &  Invariant mass of the combination of the leptons, jet, and \MET \\
Centrality$(\mathrm{j}\ell\ell)$         &  Centrality of jet and leptons, defined as ratio of transverse to total energy \\
$\HT(\text{leptons})$/\HT                &  Ratio of scalar sum of \pt of the leptons to the \HT of full system \\
$\pt(\mathrm{j}\ell\ell)$                &  Vector sum of \pt of jet and leptons \\
\MET                                     &  Missing transverse energy in the event \\
\end{scotch}
}
\end{table*}

\begin{table*}[hb]
\topcaption {Contributions to the systematic uncertainty in the measured cross section.  The values are estimated by fixing each source one at a time and evaluating the change in the measured cross section uncertainty.  Systematic uncertainties apply to all processes unless specifically noted.}
\label{tab:syst}
\centering
\resizebox{\textwidth}{!}{
\begin{scotch}{ l  c  c  l }
Systematic uncertainty & $\Delta \sigma$ (pb) & $\Delta \sigma/\sigma$  & Notes \\
\hline
ME/PS matching thresholds                     & 3.3 & 14\%   & Matching threshold $2\times$ and $1/2\times$ nominal 20\GeV value in \ttbar simulation \\
Renormalization/factorization scale           & 2.9 & 12\%   & Scale value $2\times$ and $1/2\times$ nominal value of $\mtop^2 + \sum\pt^2$ in \ttbar and \tW simulation\\
Top-quark mass                                & 2.2 & 9\%    & \mtop varied in \tW and \ttbar simulation by $\pm$2\GeV \\
Fit statistical                               & 1.9 & 8\%    & Remaining uncertainty in fit when all other systematic uncertainties are removed \\
Jet energy scale                              & 0.9 & 4\%    & Jet energy scale varied up/down \\
Luminosity                                    & 0.7 & 3\%    & 2.6\% uncertainty in the measured luminosity \\
\cPZ+jets data/simulation scale factor        & 0.6 & 3\%    & Varying scale factors used for correcting \cPZ+jets \MET simulation\\
\tW DR/DS scheme                              & 0.5 & 2\%    & Difference between DR and DS scheme used for defining \tW signal \\
\ttbar cross section                          & 0.4 & 2\%    & Uncertainty in the cross section of \ttbar production\\
Lepton identification                         & 0.4 & 2\%    & Uncertainty in scale factors for lepton efficiencies between data/simulation\\
PDF                                           & 0.4 & 2\%    & From choice of PDF\\
Jet energy resolution                         & 0.2 & 1\%    & Energy resolution for jets varied up/down \\
$\cPqb$-tagging data/simulation scale factor  & 0.2 & $<$1\% & Variations in scale factors\\
\ttbar spin correlations                      & 0.1 & $<$1\% & Difference between \ttbar simulation with/without spin correlations\\
Pileup                                        & 0.1 & $<$1\% & Varying effect of pileup\\
Top-quark \pt reweighting                     & 0.1 & $<$1\% & Uncertainty due to differences in top quark \pt between data and \ttbar simulation \\
\MET modeling                                 & 0.1 & $<$1\% & Uncertainty in amount of unclustered \MET \\
Lepton energy scale                           & 0.1 & $<$1\% & Uncertainty in energy of leptons\\
\hline
Total & 5.5 & 24\% \\
\end{scotch}
}
\end{table*}

\begin{figure*}
  \centering
    \includegraphics[width=0.3\textwidth]{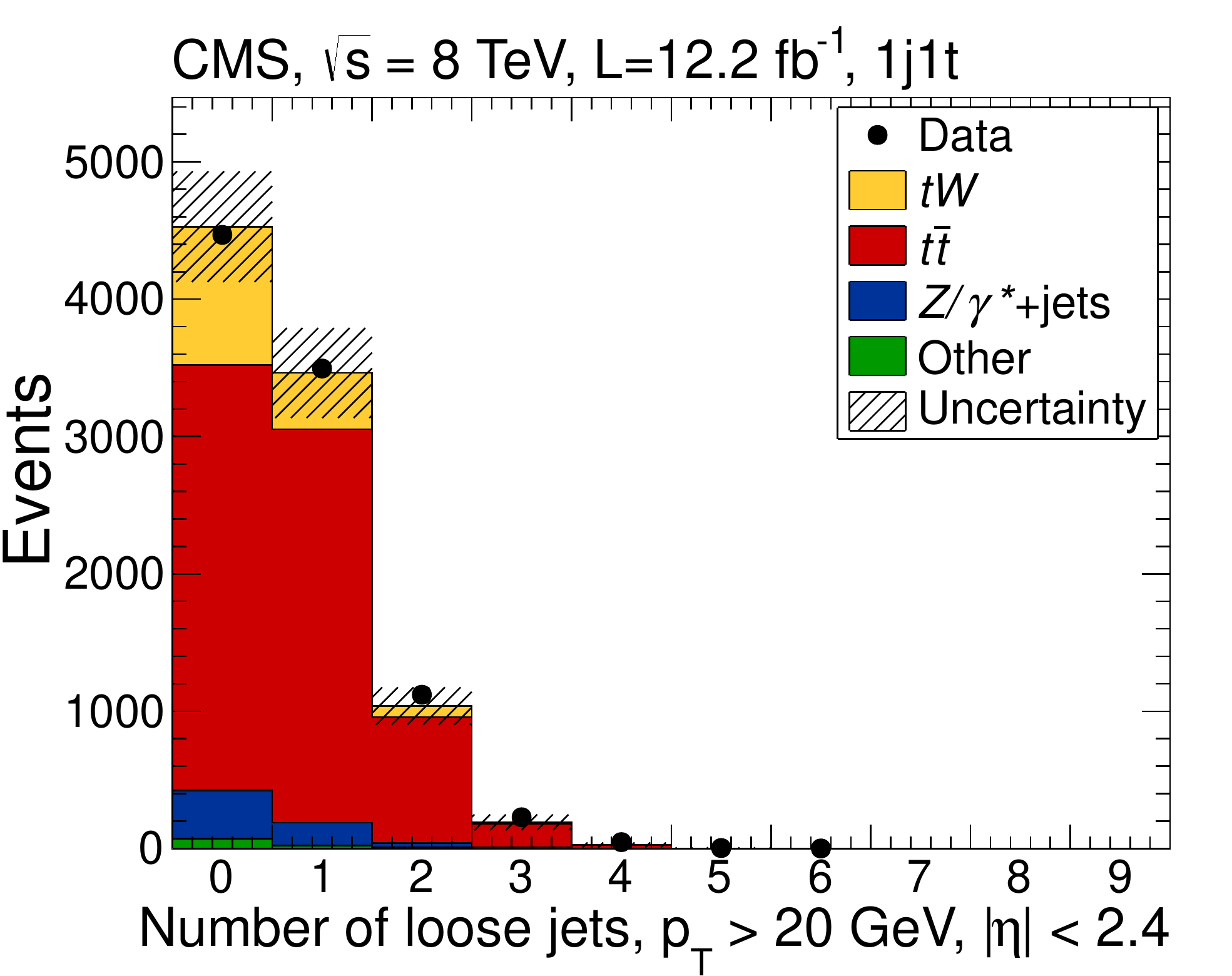}
    \includegraphics[width=0.3\textwidth]{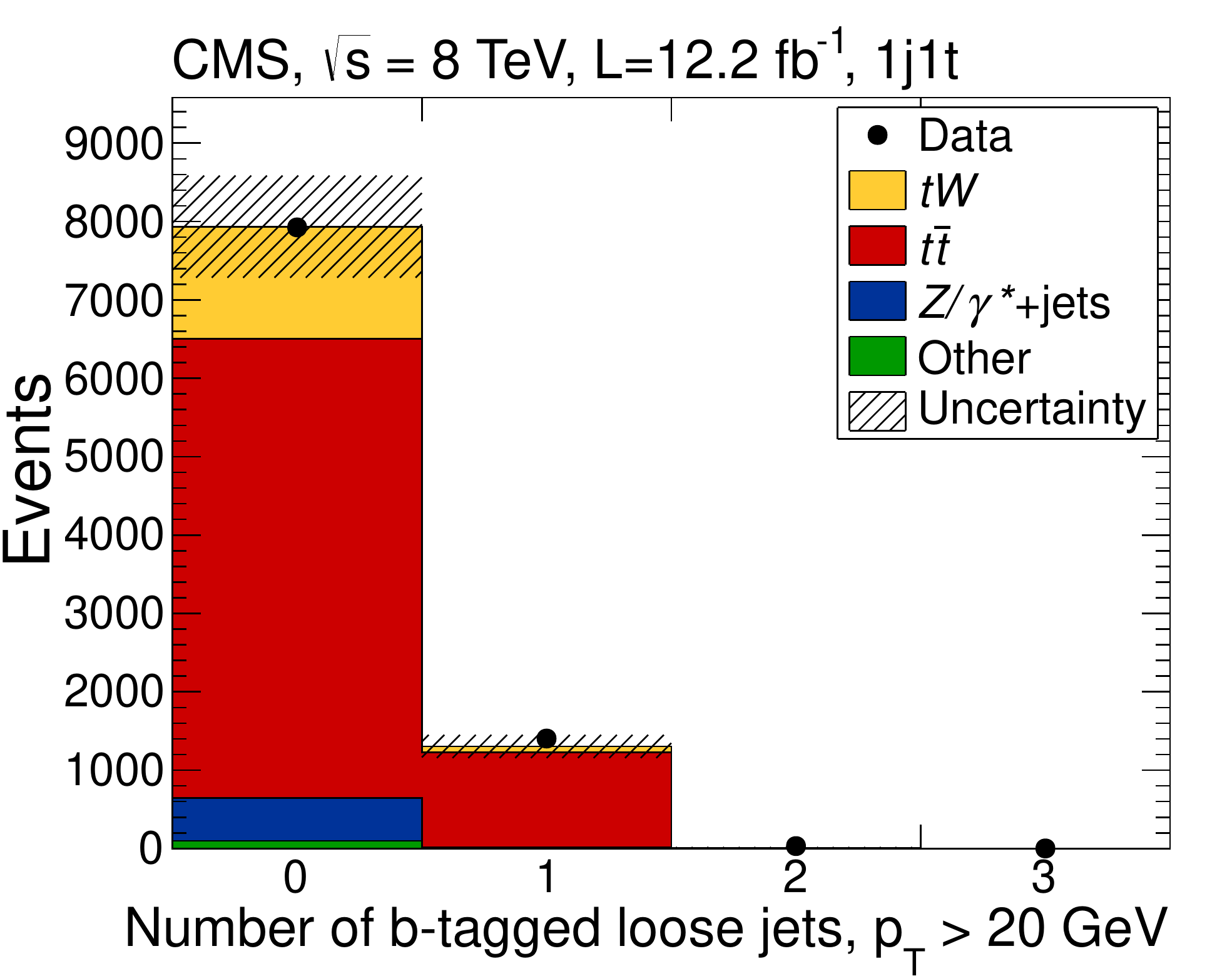}
    \includegraphics[width=0.3\textwidth]{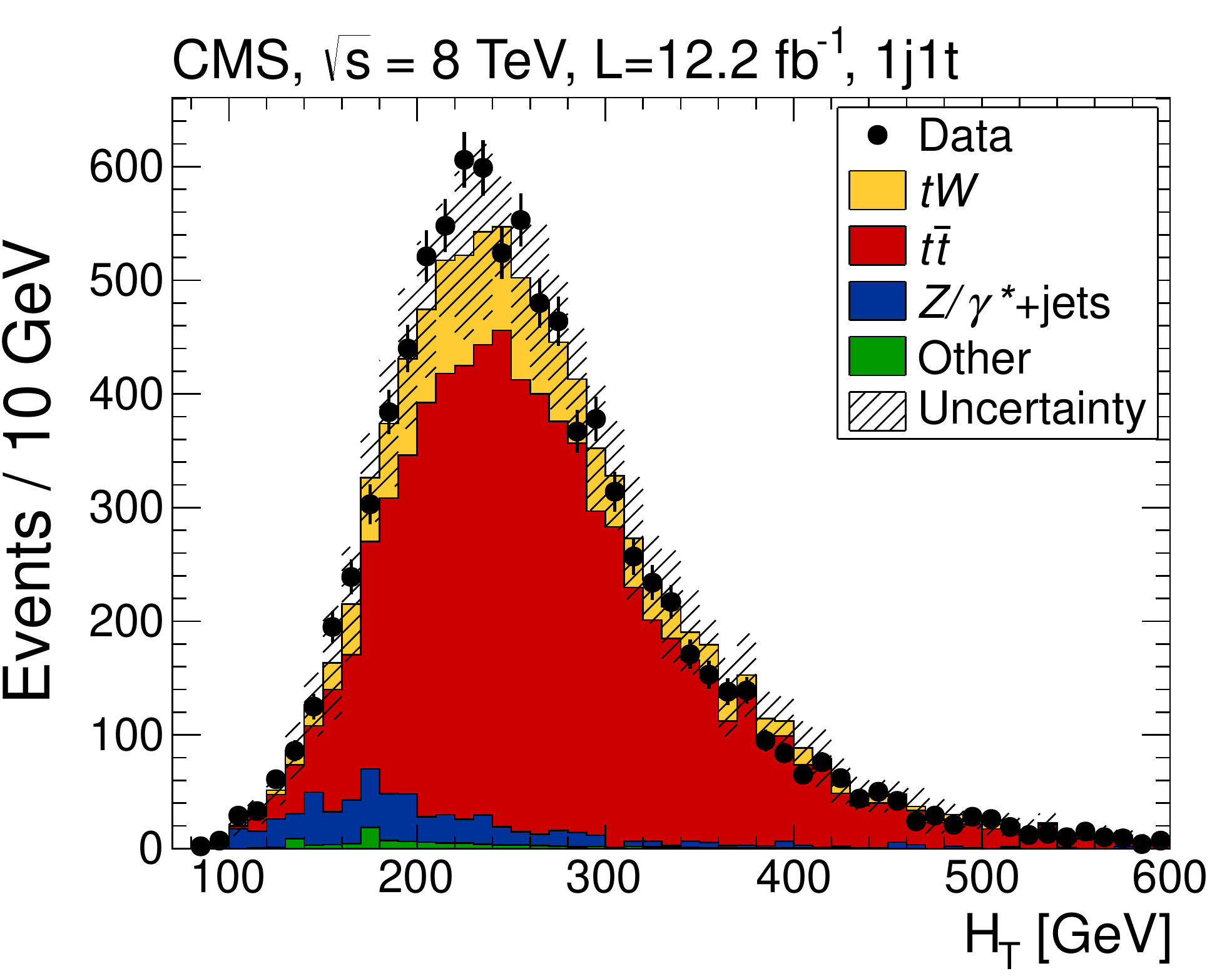} \\
    \includegraphics[width=0.3\textwidth]{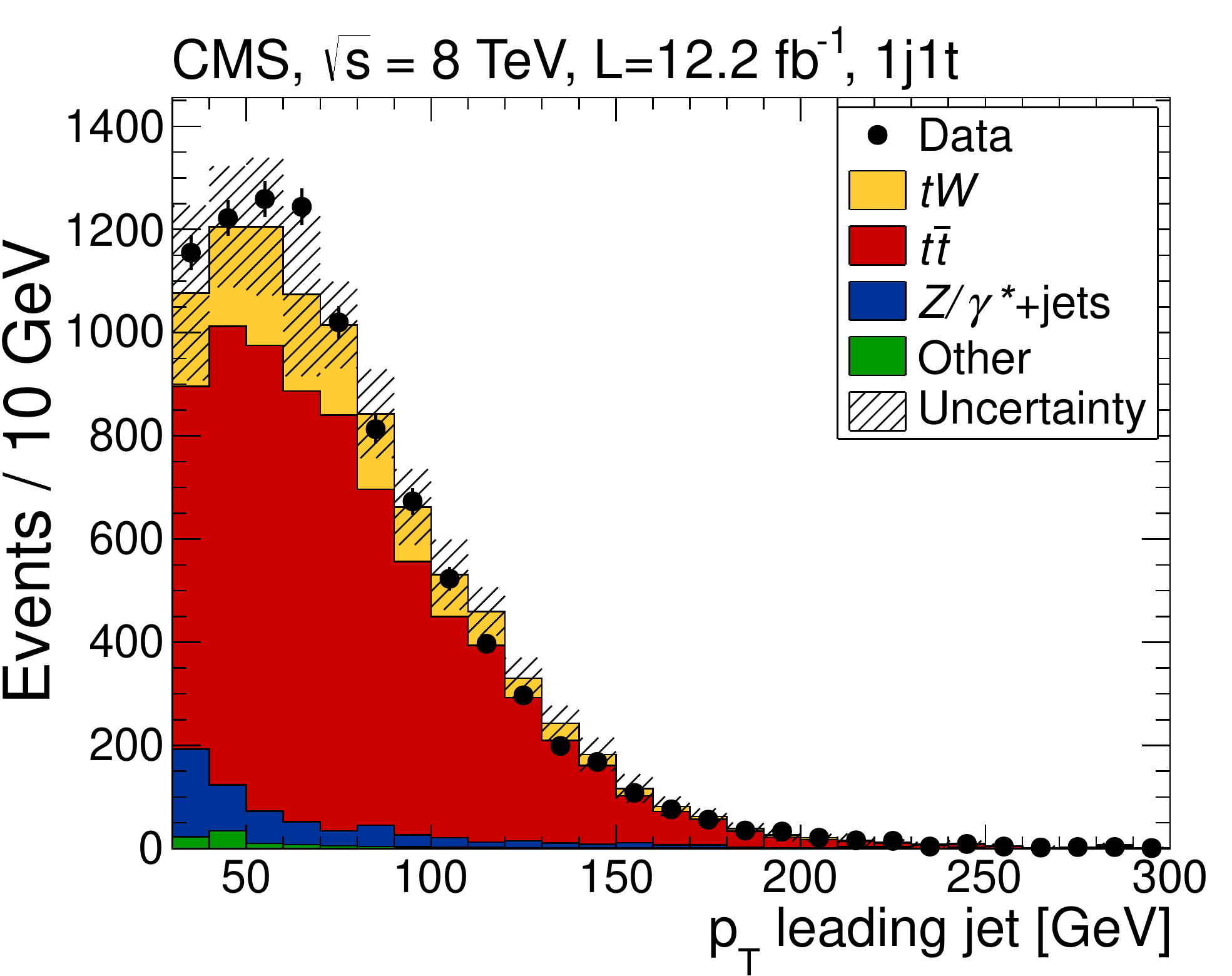}
    \includegraphics[width=0.3\textwidth]{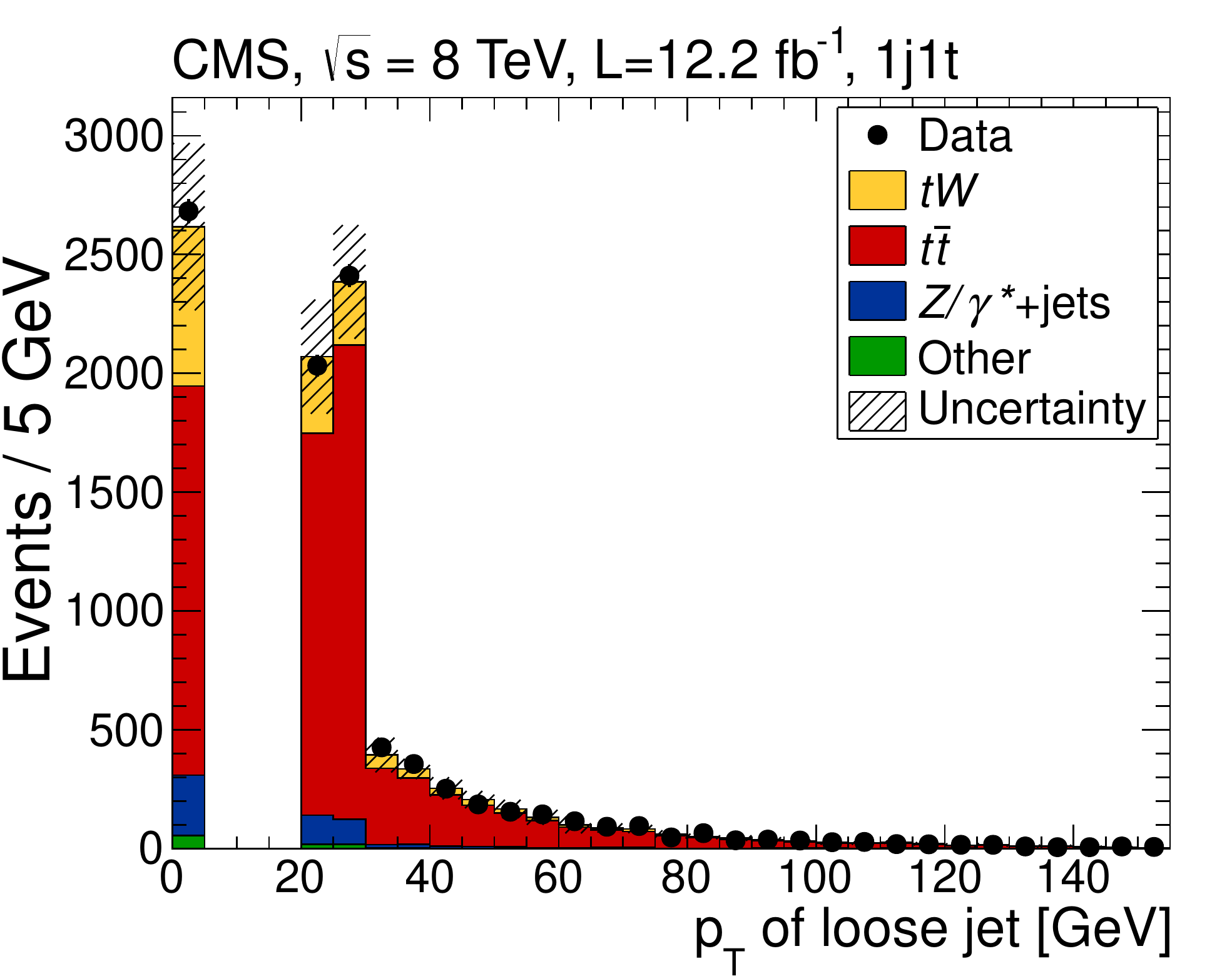}
    \includegraphics[width=0.3\textwidth]{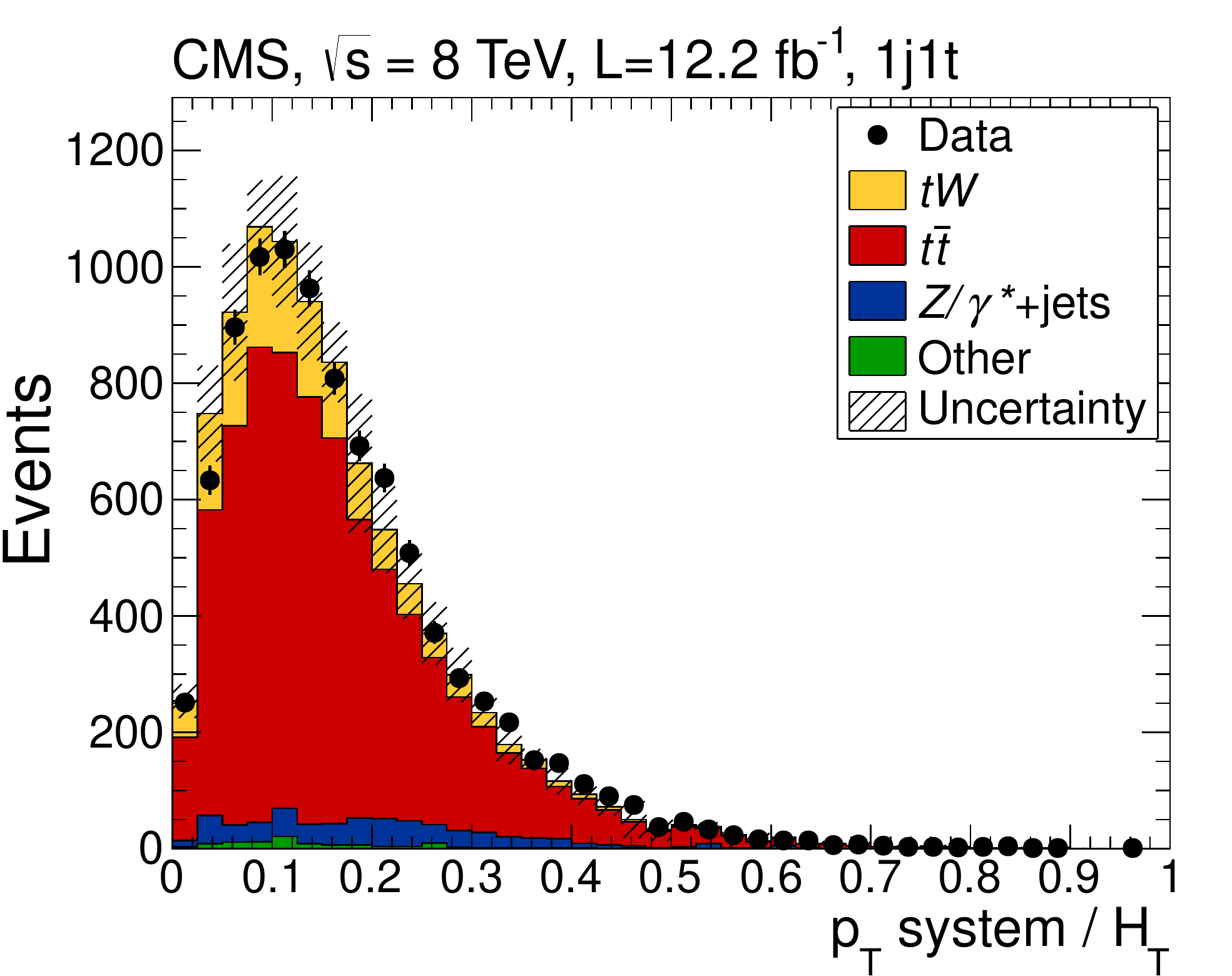} \\
    \includegraphics[width=0.3\textwidth]{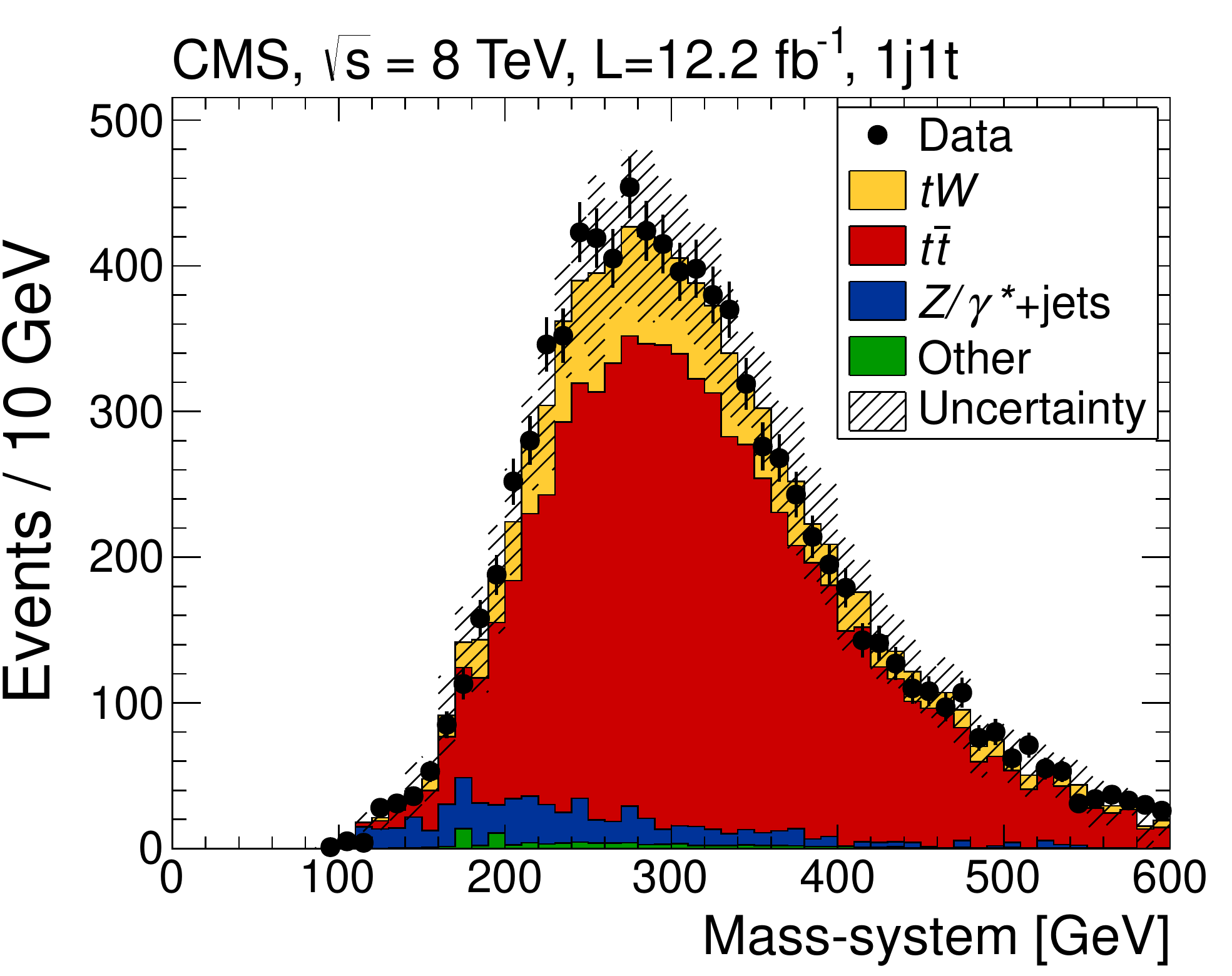}
    \includegraphics[width=0.3\textwidth]{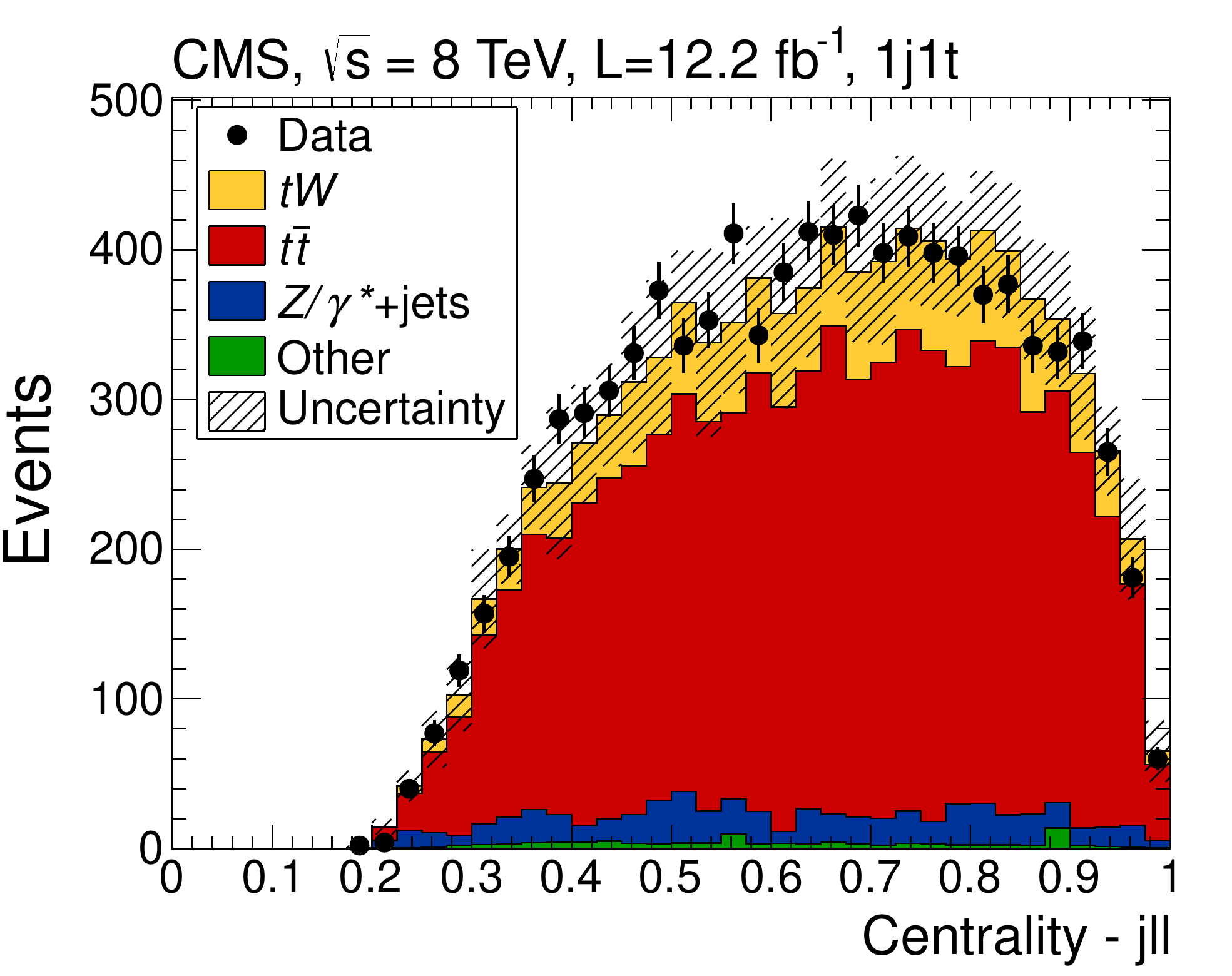}
    \includegraphics[width=0.3\textwidth]{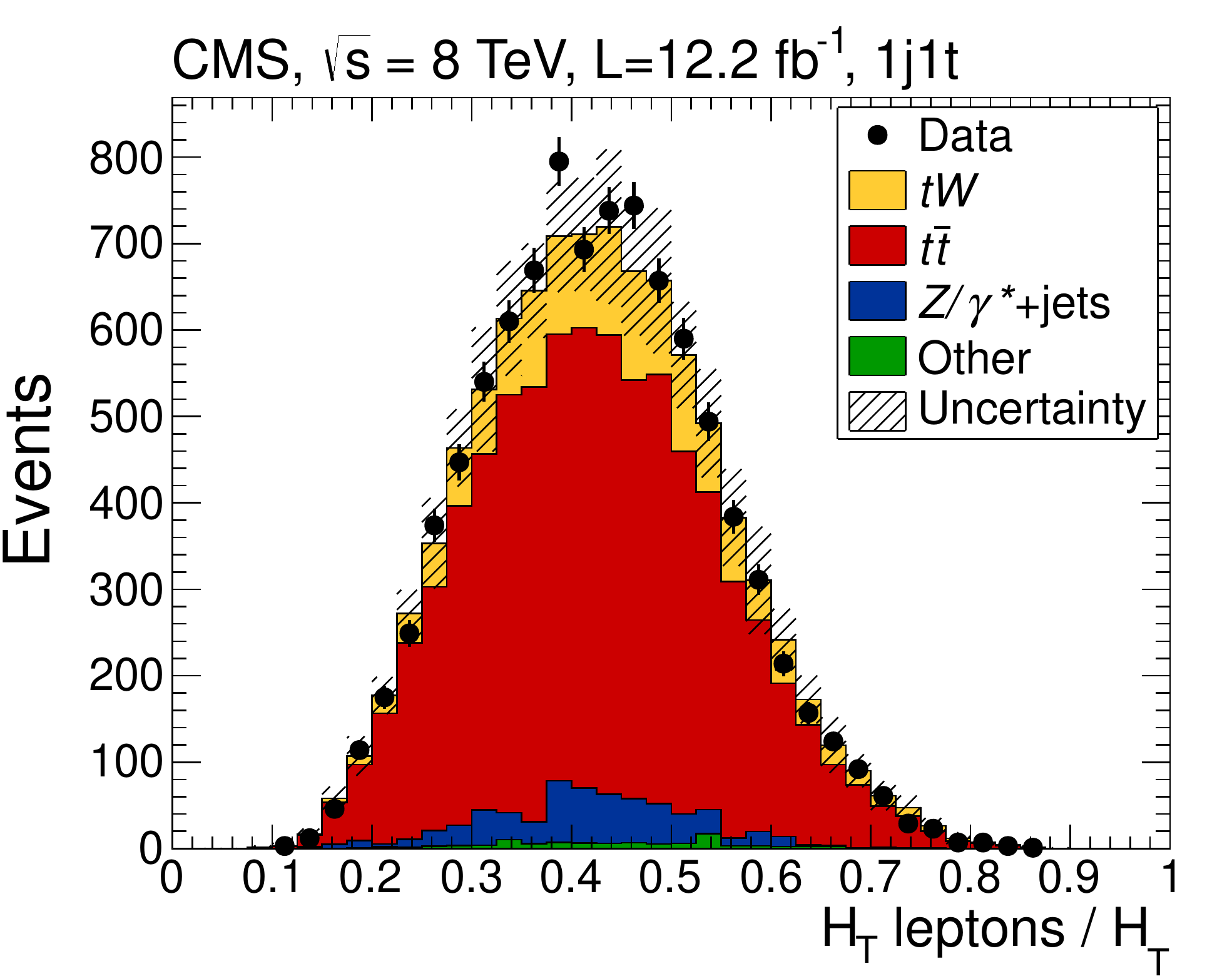} \\
    \includegraphics[width=0.3\textwidth]{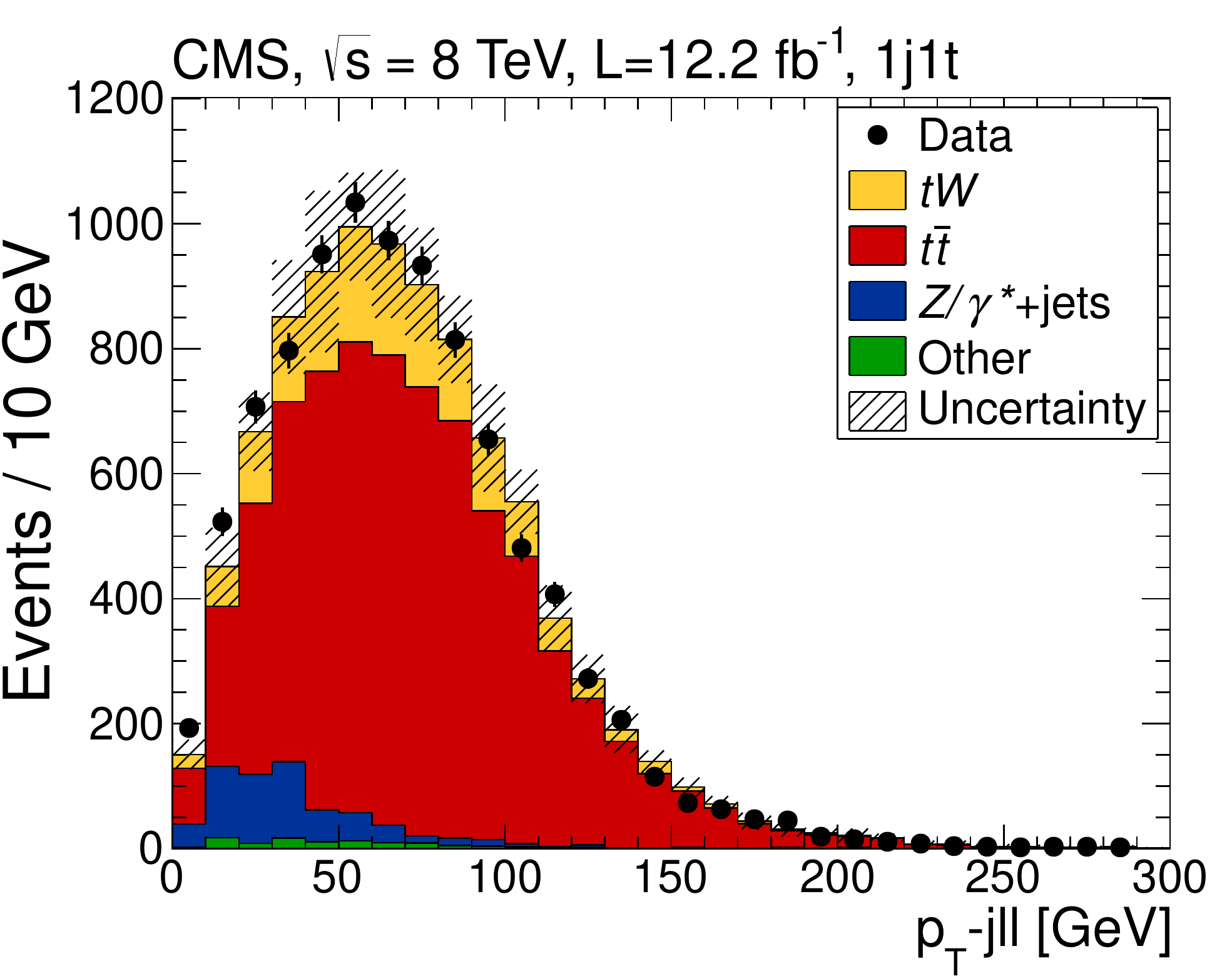}
    \includegraphics[width=0.3\textwidth]{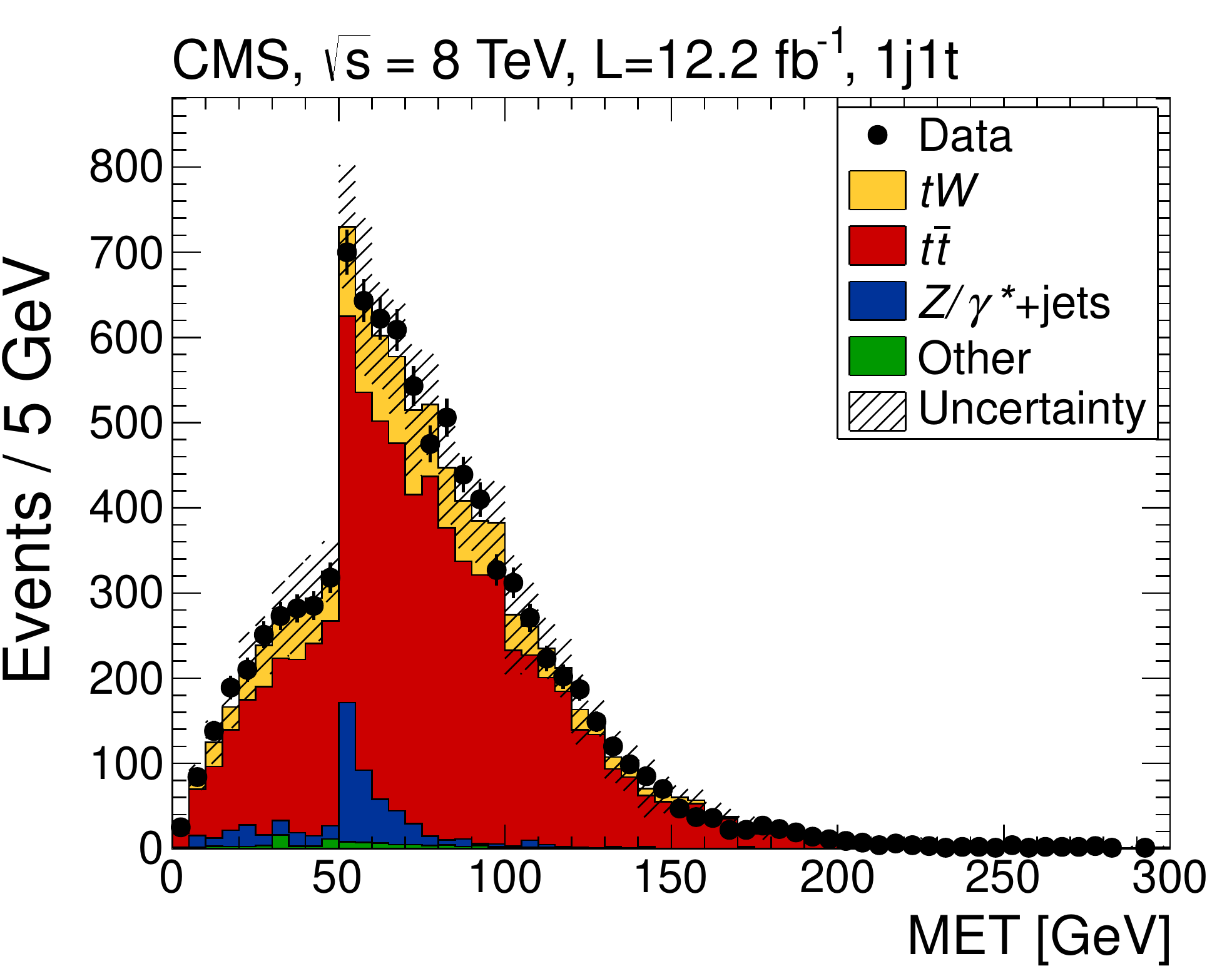}
    \caption{\label{fig:inputVariables}  The distribution of all additional input variables used in the analysis, in the signal region (1j1t) for all final states combined.  Shown are data (points) and simulation (histogram). The hatched band represents the combined effect of all sources of systematic uncertainty.}
\end{figure*}

\begin{figure*}
  \centering
    \includegraphics[width=0.3\textwidth]{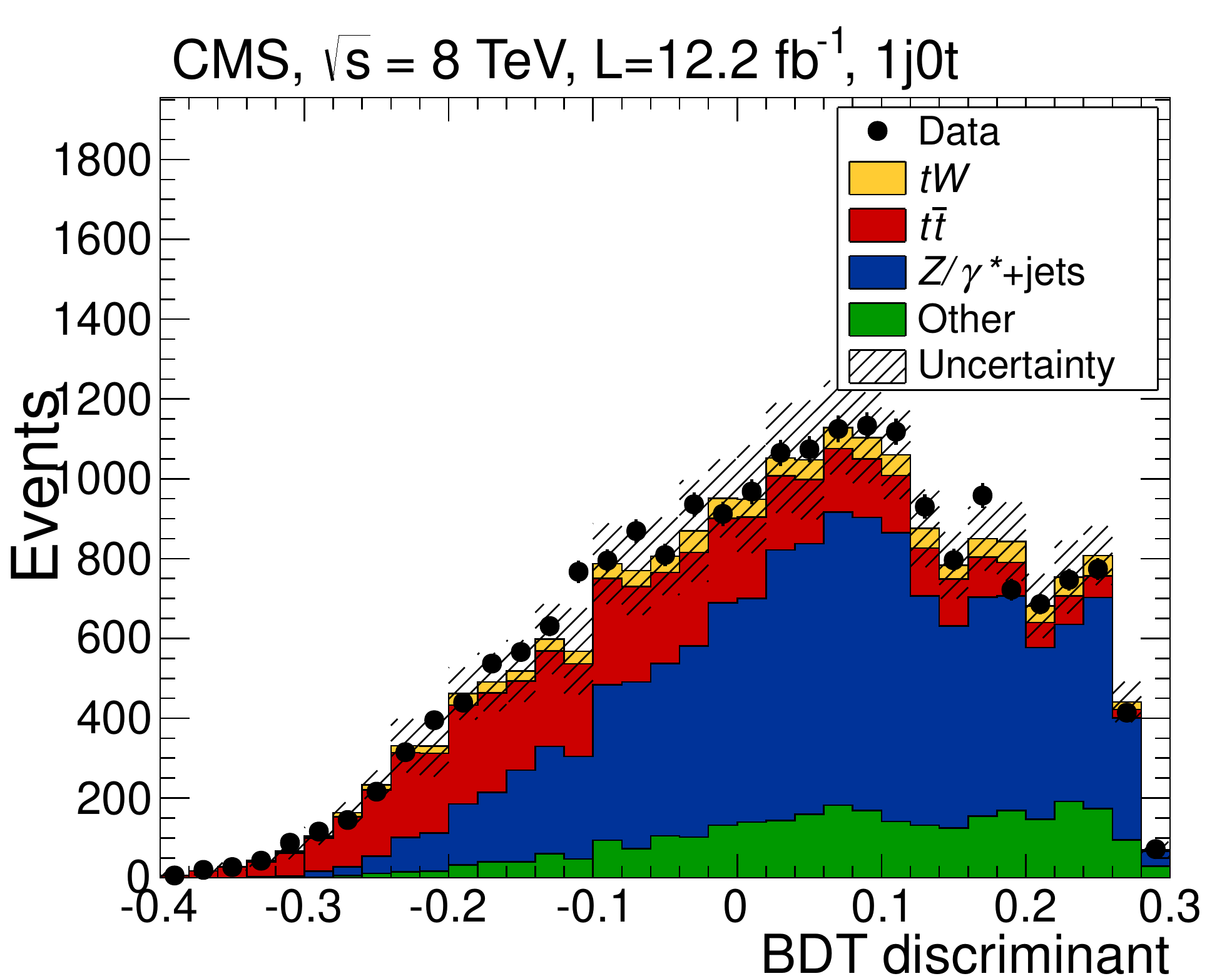}
    \includegraphics[width=0.3\textwidth]{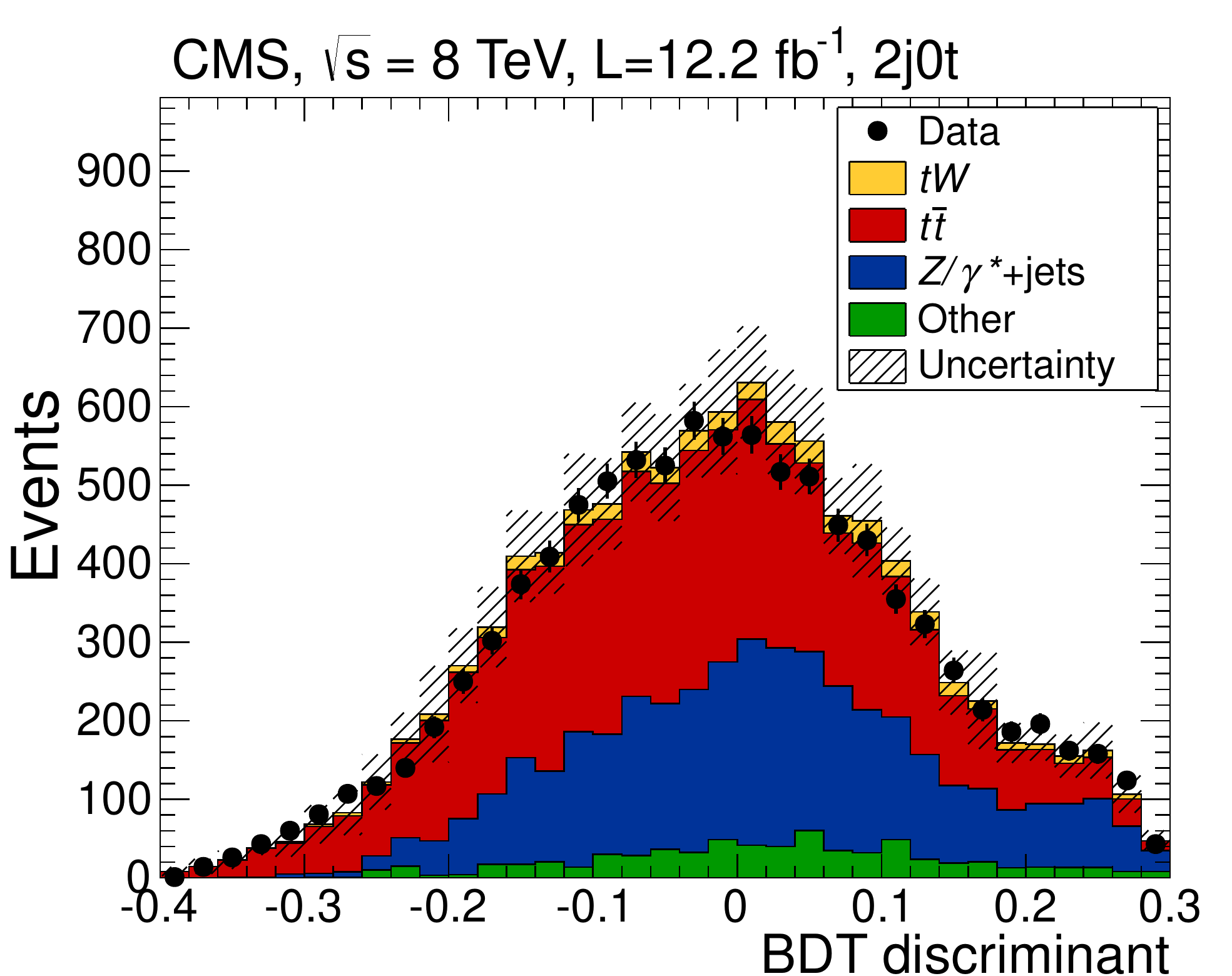}
    \caption{\label{fig:bdt0jets}  The BDT discriminant, in two control regions (1j0t and 2j0t) used for validating the shape of the discriminant in the \DY and \ttbar backgrounds. Shown are data (points) and simulation (histogram). The hatched band represents the combined effect of all sources of systematic uncertainty.}
\end{figure*}

\begin{table*}[th]
\topcaption{Event yields in the signal and control regions in the $\Pe\Pe$ final state. Yields from simulation are shown with statistical (first) and systematic (second) uncertainties.}
\label{tab:eventCountsee}
\centering
\setlength{\extrarowheight}{0.5ex}
\begin{scotch}{lccc}
$\Pe\Pe$         &  1j1t                  & 2j1t                    & 2j2t \\
\hline
\tW              & $230  \pm 10 \pm 30 $ & $130  \pm 10 \pm 20 $ & $40   \pm 5  \pm 10 $ \\
\ttbar           & $1170 \pm 30 \pm 160$ & $2140 \pm 30 \pm 230$ & $1230 \pm 30 \pm 180$ \\
\DY, other       & $170  \pm 20 \pm 30 $ & $110  \pm 10 \pm 20 $ & $10   \pm 3  \pm 6  $ \\
\hline
Tot. sim.        & $1560 \pm 30 \pm 190$ & $2380 \pm 40 \pm 250$ & $1280 \pm 30 \pm 180$ \\
Data             & 1618 & 2395 & 1318 \\
\end{scotch}
\end{table*}

\begin{table*}[th]
\topcaption{Event yields in the signal and control regions in the $\Pe\mu$ final state. Yields from simulation are shown with statistical (first) and systematic (second) uncertainties.}
\label{tab:eventCountsemu}
\centering
\setlength{\extrarowheight}{0.5ex}
\begin{scotch}{lccc}
$\Pe\mu$         &  1j1t                  & 2j1t                    & 2j2t \\
\hline
\tW              & $990  \pm 20 \pm 90  $& $500  \pm 10 \pm 50 $ & $140  \pm 10 \pm 20  $\\
\ttbar           & $4430 \pm 50 \pm 580 $& $8190 \pm 70 \pm 850$ & $4880 \pm 50 \pm 610 $\\
\DY, other       & $280  \pm 20 \pm 40  $& $130  \pm 20 \pm 30 $ & $11   \pm 3  \pm 5   $\\
\hline
Tot. sim.        & $5700 \pm 60 \pm 650 $& $8820 \pm 70 \pm 900$ & $5030 \pm 50 \pm 620 $\\
Data             & 5802 & 8324 & 4732 \\
\end{scotch}
\end{table*}

\begin{table*}[th]
\topcaption{Event yields in the signal and control regions in the $\mu\mu$ final state. Yields from simulation are shown with statistical (first) and systematic (second) uncertainties.}
\label{tab:eventCountsmumu}
\centering
\setlength{\extrarowheight}{0.5ex}
\begin{scotch}{lccc}
$\mu\mu$         &  1j1t                  & 2j1t                    & 2j2t \\
\hline
\tW              & $280  \pm 10 \pm  30$ & $170  \pm 10 \pm 20 $ & $50   \pm 5  \pm 10 $ \\
\ttbar           & $1500 \pm 30 \pm 190$ & $2580 \pm 40 \pm 270$ & $1540 \pm 30 \pm 210$ \\
\DY, other       & $220  \pm 20 \pm  50$ & $130  \pm 20 \pm 30 $ & $16   \pm 5  \pm 7  $ \\
\hline
Tot. sim.        & $2000 \pm 40 \pm 230$ & $2880 \pm 40 \pm 290$ & $1600 \pm 30 \pm 230$ \\
Data             & 1933 & 2760 & 1565 \\
\end{scotch}
\end{table*}

}

\cleardoublepage \section{The CMS Collaboration \label{app:collab}}\begin{sloppypar}\hyphenpenalty=5000\widowpenalty=500\clubpenalty=5000\textbf{Yerevan Physics Institute,  Yerevan,  Armenia}\\*[0pt]
S.~Chatrchyan, V.~Khachatryan, A.M.~Sirunyan, A.~Tumasyan
\vskip\cmsinstskip
\textbf{Institut f\"{u}r Hochenergiephysik der OeAW,  Wien,  Austria}\\*[0pt]
W.~Adam, T.~Bergauer, M.~Dragicevic, J.~Er\"{o}, C.~Fabjan\cmsAuthorMark{1}, M.~Friedl, R.~Fr\"{u}hwirth\cmsAuthorMark{1}, V.M.~Ghete, C.~Hartl, N.~H\"{o}rmann, J.~Hrubec, M.~Jeitler\cmsAuthorMark{1}, W.~Kiesenhofer, V.~Kn\"{u}nz, M.~Krammer\cmsAuthorMark{1}, I.~Kr\"{a}tschmer, D.~Liko, I.~Mikulec, D.~Rabady\cmsAuthorMark{2}, B.~Rahbaran, H.~Rohringer, R.~Sch\"{o}fbeck, J.~Strauss, A.~Taurok, W.~Treberer-Treberspurg, W.~Waltenberger, C.-E.~Wulz\cmsAuthorMark{1}
% ...
\vskip\cmsinstskip
% ...
1:~~Also at Vienna University of Technology, Vienna, Austria\\
2:~~Also at CERN, European Organization for Nuclear Research, Geneva, Switzerland\\
% ...

\end{sloppypar}
\end{document}